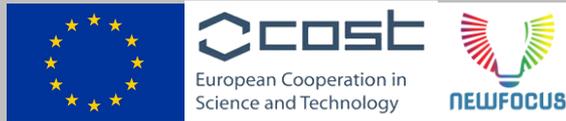

**2ⁿᵈ White Paper**

https://www.newfocus-cost.eu/action/

**May 2023**

# EU COST ACTION ON FUTURE GENERATION OPTICAL WIRELESS COMMUNICATION TECHNOLOGIES

## NEWFOCUS CA19111

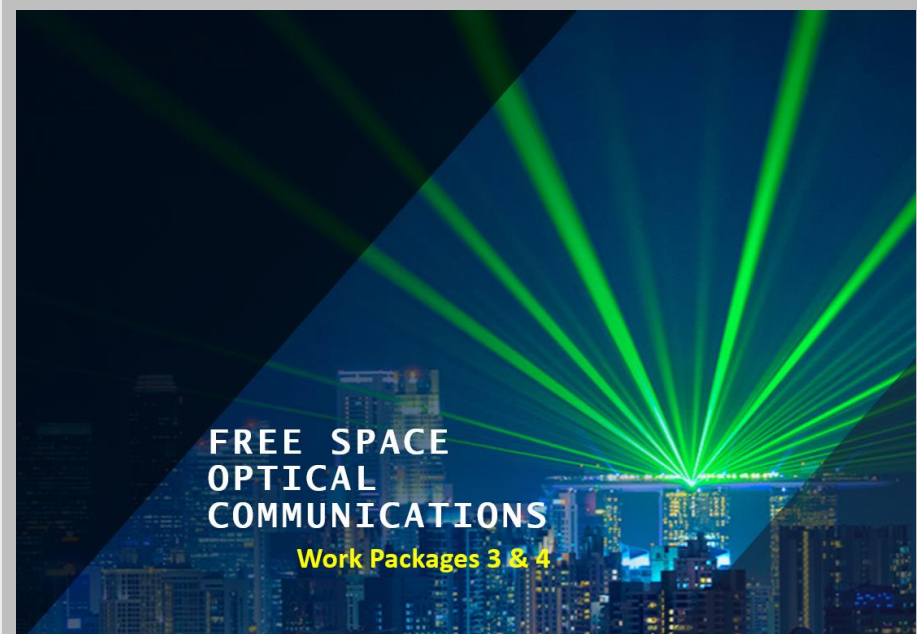

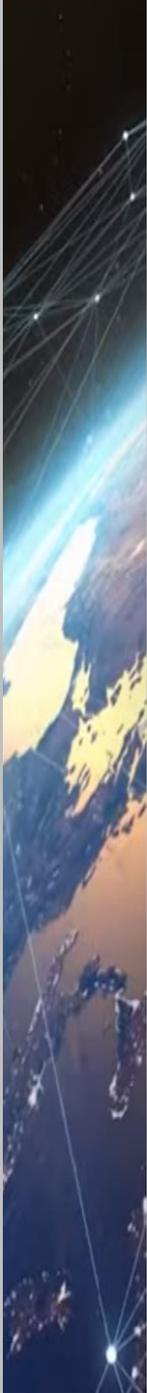

**Contents**

# INTRODUCTION

**NEWFOCUS** is an EU COST Action targeted at exploring radical solutions that could influence the design of future wireless networks[1]. The project aims to address some of the challenges associated with **optical wireless communication (OWC)** and to establish it as a complementary technology to the radio frequency (RF)-based wireless systems in order to meet the demanding requirements of the fifth generation (5G) and the future sixth generation (6G) backhaul and access networks. Only 6G will be able to widely serve the exponential growth in connected devices (i.e., more than 500 billion) in 2030, real-time holographic communication, future virtual reality, etc. Space is emerging as the new frontier in 5 and 6G and beyond communication networks, where it offers high-speed wireless coverage to remote areas both in lands and sees. This activity is supported by the recent development of low-altitude Earth orbit satellite mega-constellations[2]. The focus of this 2nd White Paper is on the use of OWC as an enabling technology for medium- and long-range links for deployment in (*i*) smart-cities and intelligent transportation systems; (*ii*) first- and last-mile access and backhaul/fronthaul wireless networks; (*iii*) hybrid free-space optics/RF adaptive wireless connections; (*iv*) space-to-ground, inter-satellite, ground-to-air, and air-to-air communications; and (*v*) underwater communications.

In this **White Paper**, we have eight contributions. The first article on "Optical Wireless Communications – Medium to Long Range Applications" by *Z. Ghassemlooy et al*, gives an overview of the OWC application, whereas the next four contributions on "Free Space Optical Communication in the Mid-IR for Future Long-Range Terrestrial and Space Applications" by *X. Pang and C. Sirtori*, "Review of Low-Earth Orbit Satellite Quantum Key Distribution", and "Perspectives for Global-scale Quantum Key Distribution via Uplink to Geostationary Satellites" by *D. Orsucci, A. Shrestha and F. Moll*, "Wavelength Division Multiplexing Free Space Optical Links" by *G. Cossu et al*, and are focused on FSO links for satellite communications, and "Review of Hybrid Optical-Radio Inter-Satellite Links in 6G NTN Including Quantum Security" by *J. Bas and M. Amay*. The next contribution is on "Highly-Sensitive SPAD-Based Optical Wireless Communication" by *S. Huang et al*. Three contributions on "On Signalling and Energy Efficiency of Visible Light Communication Systems" by *T. Gutema and W. Popoola*, "Practical Implementation of Outdoor Optical Camera Communication Systems with Simultaneous Video and Data Acquisition" by *V. Matus et al*, and "Physical Layer Security in Visible Light Communications" by - *E. Panayirci, P. D. Diamantoulakis and H. Haas*, are dedicated to the visible light communication and optical camera communication. The last contribution is on "Optical Wireless Communication Based Smart Ocean Sensor Networks for Environmental Monitoring" *by I. C. Ijeh*.

Our thanks go out to all authors for their contributions to the 2nd White Paper, which we hope will serve as a valuable resource on some of the features, challenges, and future work associated with the OWC technology.

Z. Ghassemlooy, *Vice-Chair of Newfocus*
M-A. Khalighi, *Chair of Newfocus*

---

[1] https://www.newfocus-cost.eu/action/
[2] N. Pachler, et al, "An updated comparison of four low earth orbit satellite constellation systems to provide global broadband," in Proc. IEEE Int. Conf. Commun., 2021, pp. 1–6

# Optical Wireless Communication – Medium to Long Range Applications

Zabih Ghassemlooy[1], Mohammad-Ali Khalighi[2], Stanislav Zvanovec[3], Amita Shrestha[4], Beatriz Ortega[5] and Milica Petkovic[6]

[1]*Optical Communications Research Group, Faculty of Engineering and Environment, Northumbria University, UK, z.ghassemlooy@northumbria.ac.uk*

[2]*Aix-Marseille University, CNRS, Centrale Marseille, Institut Fresnel, Marseille, France, Ali.Khalighi@fresnel.fr*

[3]*Dept. of Electromagnetic Field, Faculty of Electrical Engineering, Czech Technical University in Prague, Prague, Czech Republic, xzvanove@fel.cvut.cz*

[4]*German Aerospace Center, Germany (DE), amita.shrestha@dlr.de*

[5]*Institute of Telecommunications and Multimedia Applications, Universitat Politècnica de València, Spain, bortega@dcom.upv.es*

[6]*University of Novi Sad, Faculty of Technical Sciences, Serbia, milica.petkovic@uns.ac.rs*

## I. INTRODUCTION

It is imperative that the state-of-the-art data communication methods be used to meet the ever-increasing demand for higher data transfer bit rates at a lower cost and a lower energy consumption. The major challenges using radio frequency (RF) technology are (*i*) costly spectrum licensing; (*ii*) congested spectrum, thus moving to higher spectrum bands i.e., higher millimetre wave (up to 100 GHz) and tera Hertz (0.1 – 10 THz), which however, are limited to link spans up to 100 m due to huge path loss and severe molecular absorption; and (*iii*) vulnerability to detection, interception and jamming. Therefore, in congested and contested environments (e.g., city centres, shopping malls, airports, stadiums, etc) using RF technology may result in a poor quality of service, lack of link availability, as well as providing opportunities for eavesdropping. In fact, the common approach in RF communication systems rely upon broadcasting data widely over a pre-determined frequency band and encrypting it to prevent interception.

In the last 50 years, the deployment of optical fibre technology has played a critical role in the growth of the Internet, network systems, cable television and telecommunications, thus transforming our lives and the world, providing very high data rates over long transmission distances, without cross-talk and with high quality and reliability [1]. However, in many cases, it is impossible or impractical to lay down fibre optic cables in rural and urban areas due to high costs. As part of global networks, optical wireless transmission systems (e.g., based on **optical wireless communication (OWC)**) are the best option to complement both RF and optical fibre technologies, even though they are not intended to replace them. While civilian networks have been using RF, OWC, and hybrid RF-OWC technologies for several applications, they will continue to do so in the future. However, they are moving toward using **free space optical (FSO)** communication for a range of transmission range, data rates, and applications. FSO presents a new paradigm in communication systems as a complementary wireless technology to the RF-based systems. FSO systems typically transmit a narrow beam of light (i.e., a laser beam) between two specific points, effectively enabling a fibre optic cable in free space, and offering several interesting features as outlined in Fig. 1.

In recent years, the rapid developments and improvements in research laboratories and commercial telecommunication industry sectors have meant the components that are used in FSO systems to become smaller, compact, and more robust, thus enabling the users to bridge the technology into many diverse applications including space, terrestrial (i.e., access networks), and underwater. Every technological advancement in the last two decades, see Fig. 2, has helped make FSO communications more accessible, cheaper, faster, easier to deploy, and, ultimately, more commercially viable, as well as allowing operators to provide a bandwidth scalable system that can be used in the next generation FSO links. By merging FSO systems with the coherent technology in optical fibre communications together with adaptive optics, advanced coding/modulation and signal processing greater reach and higher capacity (i.e., tens of gigabits per second) than before can be achieved, enabling new applications in space and terrestrial links. According to Global Market estomates that the FSO sector is expected to grow from $4.4 billion in 2022 to $47.5 billion by 2027, i.e., a compound annual growth rate of about 34 % [2].

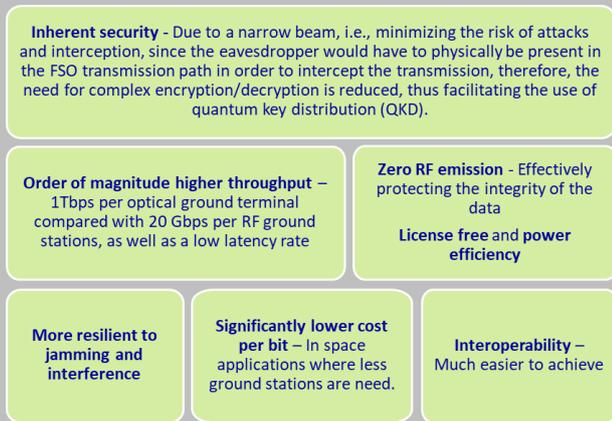

**Fig. 1.** FSO features.

FSO technology offering full duplex links with data rates of 100 Mbps to 10 Gbps and over 10 Gbps with auto-tracking and low latency (< 0.005 ms in some commercial systems, e.g., EL-10Gex from EC System) can be used in many applications including: deep-space, satellite, space-to-ground and ground-to-s[ace, terrestrial (the last mile access and cellular networks), smart environments, data centres, manufacturing, transportation, underwater, etc., see Fig. 3. The healthcare sector is one industry that stands to benefit greatly from FSO technology for medical equipment such as MRI machines and CT scanners. Few of the current key industry players in FSO communications are: Lightpointe Communication, Trimble Hungary, Wireless Excellence, fSONA Networks Corporation, Laser Light Communications, Plaintree Systems, Fog Optics, MostCom, SA Photonics, Inc., Centauri, Mynaric AG, Laser Optronics, Luma, Qinetiq, Anova Technologies, Optelix, EC Systems, and others.

*(i) Space communications*

Morgan Stanley's Space Team has estimated that $350 billion global space industry could rise to over $1 trillion by 2040, where 40 % of the space market will be dedicated to Internet services [3]. The evolving new space ecosystem includes Internet; earth observation; Internet-of-things (IoT); environmental monitoring; machine-to-machine communications; satellite-to-satellite communication; and data storage. In addition to technical advancements, the growing demand for bandwidth is also driving interest in FSO networks in space and terrestrial application since RF spectrums are congested and costly. It is becoming increasingly common for satellites to communicate via all optical links, whereas most ground-to-satellite communications use the traditional RF or microwave links. It should be noted that unlike the optical spectrum, which is licenced free, RF and microwave transmission bands requires licenses, thus leading to an increased interest in the use of high-speed (exceeding 100 Gbps) FSO systems for satellite-to-satellite, ground-to-satellite, satellite-to-high altitude platform (HAP) communications.

In addition, hybrid RF-FSO systems integrated into robust and highly scalable architectures may be used in back-haul and unmanned aerial vehicle communications (UAVs) for providing expanded coverage areas and improved system performance [4]. In hybrid-based schemes we have the options of (*i*) FSO or RF link being used either for the downlink and/or uplink transmission; and (*ii*) using the RF link as a backup when the optical link experience link outage due to the combined effects of beam misalignment, channel conditions (i.e., turbulence, fog, etc.), and mobility.

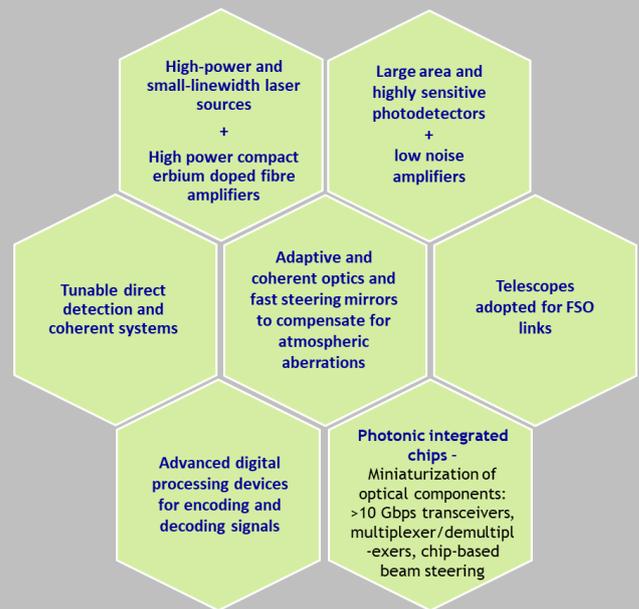

**Fig. 2.** Technological advances enabling widespread development of FSO systems.

Several companies intend to deploy mega constellations containing thousands of satellites in response to the global demand for global connectivity, which are connected through FSO links and are capable of exchanging data with each other nodes on the ground (e.g., UAVs, aircrafts, ships, and terminals), see Fig. 4. One of these satellites typically carries several optical communication terminals aimed at surrounding satellites in the same constellation. In addition to offering higher data transmission throughputs, these networks have built-in redundancy, which allows re-routing of the transmission path by bypassing malfunctioning satellites. Note that FSO-based satellite links require the laser beam from an Earth-based terminal to lock onto the satellite as it crosses the horizon, and then wait until the line-of-sight path is established prior to transmitting the information data.

In addition, there are many UAVs (i.e., high altitude platforms (HAPs) to small drones, that could be used to in terrestrial communication systems with smaller coverage areas than satellite within in 5 and 6G and beyond in application such as providing (*i*) emergency networks for disaster areas; (*ii*) extending the wireless coverage to rural areas; and (*iii*) offloading ground terminal traffic in urban areas.

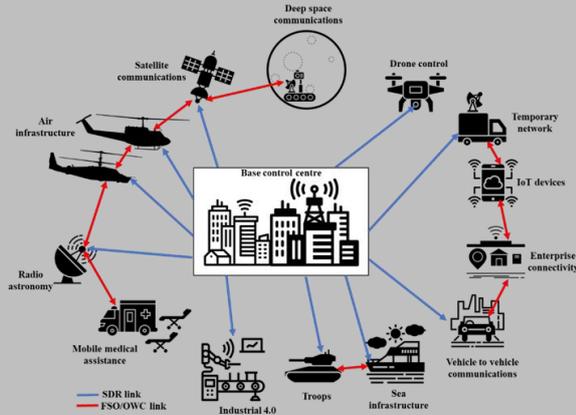

**Fig. 3.** Various potential FSO applications.

The optical feeder link between a ground station and a satellite enables a sustainable solution compared to RF technology allowing higher throughputs (e.g., 1 Tbps) at significantly lower cost/Mbps. Commercially, several satellite-to-ground links are still served with RF links, because of their low bandwidth, ease of installation, and lower cost over optical ground stations. In addition, RF offers higher availability links for users under clouds. The FSO technology can transmit encrypted data at higher rates by utilizing conventional wavelength division multiplexing techniques, as well as lending itself well to quantum-based optical techniques, where single-photon-based optical sources and detectors can be used to encrypt information and ensure its secure transmission over long distances. Utilisation of quantum key distribution allows several advantages like network flexibility, and long-term security, though several impairments must be addressed [5].

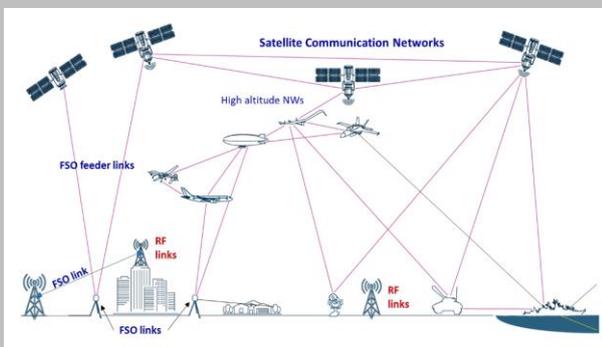

**Fig. 4.** Satellite communication networks.

The European Data Relay System (EDRS), a commercial partnership agreement between Airbus and the European Space Agency (ESA), was the first to report laser satellite communications at gigabit speeds. It is expected that as the demand for low-Earth-orbit constellations (LEOCs) increases, so will the demand for satellites, which, in turn, will lead to further optimization and commercialization of the necessary technology. Several LEOCs are currently being developed for launching in the mid-term, which will result in an increase in the number of FSO terminals in the orbit. These trends are being driven by organizations and initiatives such as the Telesat Lightspeed LEO Network (188 satellites), Rivada Space Networks (600 satellites), Rivada Space Networks (600 satellites), and Project Kuiper (3236 satellites), and SpaceX's Starlink (12,000 satellites, orbiting at an altitude of 550 km, with data rates of > 100 Mbps and an expected latency of less than 20 ms for ground terminals). The advantages of these FSO-based constellations over fixed, below-ground optical fibre communications links are that they are mobile and can be moved to wherever the need arises, such as during natural, or manmade, disasters, for example. However, common challenges for FSO links in space communications, ground-to-space, and space-to-ground communications are shown in Fig. 5.

*(ii) Last meter and last miles terrestrial communications*

The World Bank estimates that 80% of the population in developed countries have access to broadband high-speed internet compared to only 35% of the population in developing countries [6]. FSO communication offers the potential for better connection for groups that already have broadband internet access, as well as those that do not. As compared to RF wireless communication, optical communication provides bandwidth increases of 10 to 100 times. Due to labour and digging costs, the costs of setting up ground-based radio stations to receive FSO signals are also significantly lower than those associated with installing new optical fibre connections.

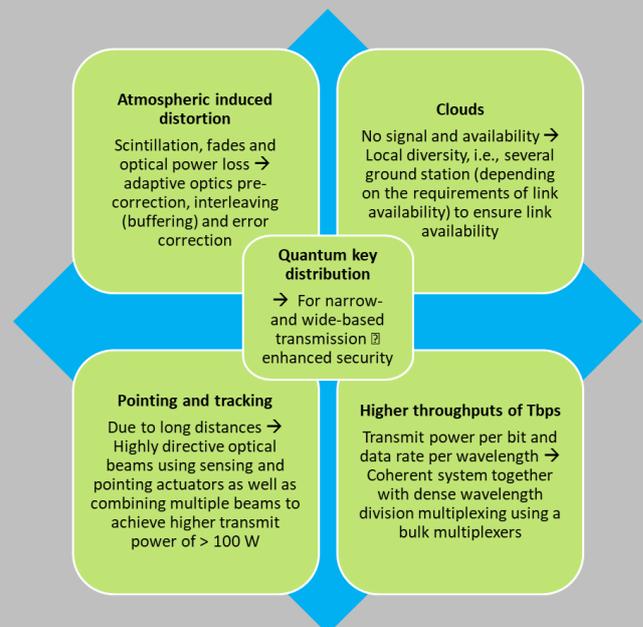

**Fig. 5.** Challenges in FSO.

There is a limit to the speed and range of current microwave and RF technologies, i.e., beyond 10 Gbps data rates using microwave technology, and higher than 80 Gbps utilizing millimetre wave technology. Additionally, optical fibre deployments are not cost-effective for rural areas with a small number of customers. As the prices of key optical components continue to decrease, terrestrial FSO systems are becoming more competitive with fibre-based systems. In recent years, there has been a significant advancement in the optical transceiver technologies. It is anticipated that many FSO systems will be installed for use in terrestrial environment, which will contribute to further advancements in this area. In general, second-tier, or lower, operators are willing to adopt a new technology if it can increase the market share at a reduced total cost of ownership. Therefore, FSO can play a crucial role compared to alternative RF-based technologies. FSO-based networks rely on point-to-point or point-to-multi-point links between a transmitter and single or multiple receivers. However, when combined with local area networks and wireless local area networks the FSO links can provide very effective solutions to many scenarios as shown in Table I.

It is also possible to deploy hybrid RF-FSO in this application scenario to always ensure link availability under all weather conditions. This can be achieved either by means of parallel transmission of RF and FSO signal or switching-over between them. Using hybrid THz-FSO technology can significantly improve the connectivity between small cells and the core network as well as overcoming the scarcity of bandwidth.

*(iii) Underwater communications*

Establishing underwater communications using wireless technologies eliminates the need to lay expensive fibres/coaxial cables across large distances to facilitate communication. There are distinct trade-offs between range and data rates in the underwater environment for all forms of communication technologies including (*i*) **RF** – Have a very limited transmission range, a limited data rate (i.e., 1 bit/Hz of the available bandwidth), and higher signal-to-noise ratio (SNR) requirements since RF signals are exponentially attenuated due to the high electrical conductivity of seawater [7]; (*ii*) **acoustic** – Have longer transmission range up to a few kilometres with a data rate typically on the order of tens of Kbps. However, they suffer from large latency (due to low propagation speed of acoustic waves), inter-symbol interference due to multiple acoustic reflections from the surface and bottom, and Doppler spread [8]. Additionally, acoustic communication is not suitable for seabed observations due to high security requirements; and (*iii*) **optical** – Have a range of data rates of few Mbps using LEDs, and potentially up to a few Gbps using laser diodes, thus enabling real-time video transmission for different applications including underwater vehicles and remote monitoring of underwater stations, etc., yet over moderated distances [9,10].

Since near-infrared wavelengths are strongly absorbed by water, blue-green wavelengths are mostly used in underwater communications in conjunction with optical filters (to suppress ambient light). Unlike terrestrial FSO communications, underwater communication is mostly not impacted by turbulence. However, there are few drawbacks with FSO in underwater environments. The communication links are highly susceptible to (*i*) beam blocking (i.e., sediment deposits or passing undersea life); (*ii*) scintillation and beam wandering due to underwater currents; (*iii*) establishing direct line of sight due to the uneven surface of the ocean floor; and (*iv*) absorption and scattering that limit the transmission range to a few hundred meters. FSO systems available in the market include Sonardyne Bluecomm™ modems, Hydromea LUMA™ modems, and Shimadzu MC optical modems with different range, operation depth, and optical transmit power levels.

**Table I.** FSO for terrestrial application.

| |
|---|
| A bridge between LAN-to-LAN connections in urban areas i.e., connecting enterprise buildings; and WLAN-to-WLAN connections - Hospitals, manufacturing sites, and campuses offering Fast Ethernet or Gbps Ethernet for many subscribers simultaneously |
| Wireless services in rural areas, where there is no physical access to high-speed Internet |
| High-speed links between the access and clint networks to backbone optical fibre core networks |
| Converged Voice-Data-Connection |
| Temporary WLAN network for events, emergency situations, etc. |
| Back-up links in critical operations |
| Plug and play for IOT with low power |
| Distributed optical (laser) powering of IoT devices |
| Optical sensing of physiological and behavioural traits |

II. CHALLENGES AND FUTURE DEVELOPMENT OF FSO TECHNOLOGY

FSO systems are expected to experience improved performance and dependability due to advances made in component fabrication and device manufacturing, coding/modulations, signal processing, adaptive optics, beam steering, etc. The use of FSO technology in some applications may be limited due to the requirement for a clear line of sight between transmitting and receiving points as well as the associated cost. The space is poised to take-off and represent a significant new market opportunity for optical communications. Therefore, there are opportunities to leverage optical communication technology for space applications. The challenges in OWC including FSO and the future works are listed in Table II.

**Table II.** Challenges and future works.

| | |
|---|---|
| 1 | ***Hermetic packaging*** - The biggest barrier to implementation of FSO links is investment in space qualification, therefore, adopting hermetic packaging could speed up this process and help to address most of the environmental challenges. |
| 2 | ***FSO with quantum encryption*** - The challenge of maintaining a communication link and combating |
| | atmospheric conditions while the transmitter and the receiver move randomly needs addressing. |
| 3 | *FSO for IoT* - The future of telecommunications will be shaped by the IoT, i.e., connecting and establishing data transmission between billions of devices, such as vehicles, smart devices, lighting fixtures, surveillance cameras, etc. In this scenario, OWC and specifically FSO technology will play a critical role, thus need more compact, less expensive, and low power devices. |
| 4 | *Standardization* - 6G will require more bandwidth, which is a challenge that all currently used technologies must meet. FSO, therefore, plays a key role here, providing operators with an additional technology capable of handling the types of bandwidth that will be required in the backhaul network. |
| 5 | *FSO with high pointing accuracy and tracking* - Which is several orders of magnitude more complex than in RF links. E.g., sub-microradian accuracy in deep space FSO links compared with the order of milliradians in the Ka band (30 GHz) RF links [11]. |
| 6 | *Software-defined radio/optic and machine learning* – For reconfiguring modules, predicting link performance, as well as switching between communication modes and sensing capabilities, thus offering unprecedented multi-functionality in wireless sensor networks. |
| 7 | *Single photon avalanche diodes technology* - Improving the optical receiver sensitivity offering high detection, high accuracy, and high SNR. |
| 8 | *Space qualified photonic integrated systems (a single chip)* – Incorporating low noise and high gain optical amplifiers, optical transceivers (> 10 Gbps), optical multiplexer/demultiplexers(ITU-grid DWDM 1.6 nm channel spacing); high power lasers (1 W or more with high stability), high power boosters (> 100 W) at the ground level; phase array and metamaterials; large area, high speed, and low noise photodetectors; coherent combining of multiple beams to combat turbulence; dynamic beam (allowing smaller lenses to be used at receivers, i.e., beam shaping and beam focusing to concentrate the transmit power on target); and optical phase arrays. |
| 9 | *Aggregated multi-carrier* - For space applications with flexibility at the optical level with optical switching for routing and optical bypass, etc. |
| 10 | *Vertical FSO networks with automatic repeat request* - Using FSO links with a wide coverage optical footprint for multiple users. |
| 11 | *Intelligent reflecting surface aided FSO links* - In the absence of a direct FSO link, it serves as a reflector to direct the incident optical beam to the receiver. |

## REFERENCES

[1] D. Wang, Q. Sui and Z. Li, "Toward universal optical performance monitoring for intelligent optical fiber communication networks," *IEEE Communications Magazine*, vol. 58, no. 9, pp. 54-59, September 2020,

[2] https://www.photonics.com/Articles/Free-Space_Optical_Communications_Soar_with_the/a68666

[3] https://www.morganstanley.com/Themes/global-space-economy

[4] S. R., S. Sharma, N. Vishwakarma, and A. S. Madhukumar, "HAPS-based relaying for integrated space-air-ground networks with hybrid FSO/RF communication: A performance analysis," *IEEE Trans. Aerosp. Electron. Syst.*, vol. 57, no. 3, pp. 1581–1599, Jun. 2021.

[5] Y. Cao, Y. Zhao, Q. Wang, J. Zhang, S. X. Ng and L. Hanzo, "The evolution of quantum key distribution networks: on the road to the qinternet," *IEEE Communications Surveys & Tutorials*, vol. 24, no. 2, pp. 839-894, Secondquarter 2022

[6] The World Bank, Connecting for inclusion: broadband access for All. World Bank. https://www.worldbank.org/en/topic/digitaldevelopment/brief/connecting-for-inclusion-broadband-access-for-all, 2019.

[7] X. Che, I. Wells, G. Dickers, P. Kear, and X. Gong X, "Re-evaluation of RF electromagnetic communication in underwater sensor networks," *IEEE Commun. Mag.*, vol. 48, pp. 143–151, 2010, doi: 10.1109/MCOM.2010.5673085

[8] B. Han, W. Zhao, Y. Zheng, et al. "Experimental demonstration of quasi-omni-directional transmitter for underwater wireless optical communication based on blue LED array and freeform lens," *Opt. Commun.*, vol. 434, pp. 184–190, 2019, doi: 10.1016/j.optcom.2018.10.037

[9] Y. Chen, L. Zhang, and Y. Ling Y, "New approach for designing an underwater free-space optical communication system," *Front. Mar. Sci*. vol. 9, 971559, 2022, doi: 10.3389/fmars.2022.971559

[10] A. Al-Halafi, H. M. Oubei, B. S. Ooi, ans B. Shihada, "Real-time video transmission over different underwater wireless optical channels using a directly modulated 520 nm laser diode," *J. Opt. Commun. Netw.*, vol. 9, pp. 826–832, 2017, doi: 10.1364/JOCN.9.000826.

[11] A. Carrasco-Casado, and R. Mata-Calvo, "Free-space optical links for space communication networks," Ch. 8, Springer Handbook of Optical Networks, Editors: B. Mukherjee, et ak, Springer, 2020

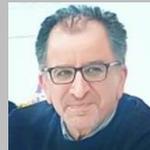

**Prof. Zabih Ghassemlooy** (Fellow, OPTICA; Fellow, IET; S. M. IEEE; CEng, BSc (Hons.) in EEE, Manchester Metropolitan Univ. (1981), MSc (1984) and PhD (1987) Manchester Univ., UK. 1987-88 Post-Doctoral Research Fellow, City Univ., UK. 1988-2004 Sheffield Hallam Univ., and 2004-14 Associate Dean Research, Faculty of Eng. & Env., Northumbria University, and is Head of Optical Communications Research Group. Research Fellow (2016-) and a Distinguished Professor (2015-) at Chinese Academy of Science. Vice-Chair of EU Cost Actions IC1101 (2011-16) and CA19111 (2020-2024) Over 980 publications (425 J. and 8 books), 115 keynote/invited talks, supervised 12 Research Fellows and 75 PhDs. Research interests: OWC, FSO, VLC, RF-OWC, software-defined networks. He is Chief Editor of British J. of Applied Science and Technology and International J. of Optics and Applications, Associate Editor of several international journals, and Co-guest Editor of several special issues OWC. Vice-Cahir of OSA Technical Group of Optics in Digital Systems (2018-); Chair IEEE Student Branch at Northumbria University, Newcastle (2019-). 2004-06 was the IEEE UK/IR Communications Chapter Secretary, Vice-Chai (2006-2008), Chair (2008-2011), and Chair IET Northumbria Network (2011-2015).

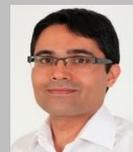

**Dr. Mohammad-Ali Khalighi** is Associate Professor with École Centrale Méditerranée, Marseille, France, and head of "Optical Communications for IoT" group at Fresnel Institute research lab. He is currently serving as Action Chair for the COST Action CA19111 NEWFOCUS (European Network on Future Generation Optical Wireless Communication Technologies, 2020-2024), and also served as Project Coordinator for the H2020 ITN MSCA VisIoN project (Visible-light-based Interoperability and Networking, 2017-2022). He is also involved in a newly-funded Horizon Europe DN-MSCA project OWIN6G (Optical and wireless sensors networks for 6G scenarios). He has co-edited the book "Visible Light Communications: Theory and Applications" (CRC Press, 2017) and was the co-recipient of the 2019 Best Survey Paper Award of the IEEE Communications Society. He is also serving as Editor-at-Large for the IEEE Transactions on Communications, and served as Associate Editor for the IET Electronics Letters as well as Guest Editor for the IEEE Open Journal of the Communications Society and Elsevier Optik journal.

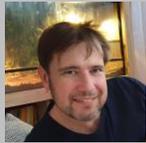**Prof. Stanislav Zvanovec** (Senior Member, IEEE) received the M.Sc. and Ph.D. degrees from the Faculty of Electrical Engineering, Czech Technical University (CTU) in Prague, in 2002 and 2006, respectively. He is currently works as a Full Professor, the Deputy Head of the Department of Electromagnetic Field, and the Chairperson of Ph.D. Branch with CTU. He leads Wireless and Fiber Optics team (optics.elmg.org). His current research interests include free space optical and fiber optical systems, visible light communications, OLED, RF over optics, and electromagnetic wave propagation issues for millimeter wave band. He is the author of two books (and coauthor of the recent book Visible Light Communications: Theory and Applications), several book chapters and more than 300 journal articles and conference papers.

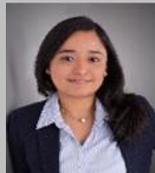**Amita Shrestha** from German Aerospace Center, received her bachelor's degree in Electronics Engineering and communications from Kathmandu University, Nepal in 2006. She received her masters' degree in Communications Systems and Electronics form Jacobs University, Bremen in 2009. Since 2010 she is working in the Institute of Communications and Navigation in DLR focusing on Free space Optical Communications. In DLR, she has been involved in the development of real-time tracking software of Institute's optical ground stations, and its operation during several satellite and aircraft downlink experiments like OPALS, SOTA, DoDfast, VABENE etc. Currently, she is actively involved in the standardization of optical links in CCSDS (The Consultative Committee for Space Data Systems) community and several other projects related to free space optical classical and quantum communications. Additionally, she is chairing one of the working groups in COST Action CA19111 NEWFOCUS, 2020-2024) focusing on long range free space optical communication links.

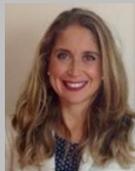**Prof. Beatriz Ortega** (Senior Member, IEEE) received the M.Sc. degree in Physics in 1995 from the Universidad de Valencia, and the Ph.D. in Telecommunications Engineering in 1999 from the Universidad Politécnica de Valencia. She currently works at the Departamento de Comunicaciones from the Universitat Politècnica de València, where she holds a Full Professorship since 2009 and collaborates as a group leader in the Photonics Research Labs in the Institute of Telecommunications and Multimedia Applications. She has published more than 200 papers and conference contributions in fibre Bragg gratings, microwave photonics and optical networks. She has got several patents and is also a co-founder of EPHOOX company. She has participated in a large number of European Networks of Excellence and R&D projects and other national ones. Her main research is currently focused on optical networks, wireless communications and microwave photonic systems and applications.

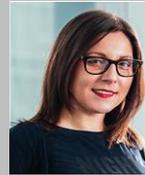**Dr. Milica Petkovic** (Member, IEEE) received her M.Sc. and Ph.D. degrees in electrical engineering from the Faculty of Electronic Engineering, University of Nis, Serbia, in 2010, and 2016, respectively. Currently, she is an Assistant Professor at Faculty of Technical Science, University of Novi Sad, Serbia. Her research interests are in the broad area of Digital Communications Systems and Signal processing, with emphasis on Optical Wireless Communications.

# Free Space Optical Communication in the Mid-IR for Future Long-range Terrestrial and Space Applications

Xiaodan Pang[1] and Carlo Sirtori[2]
[1]KTH Royal Institute of Technology, Sweden, xiaodan@kth.se
[2]Laboratoire de Physique de l'École normale supérieure, France, carlo.sirtori@ens.fr

## I. Introduction

As the data rate demand for next-generation wireless communication technologies (sixth generation - 6G) increases, an all-spectrum communication paradigm has been proposed to facilitate cooperative free-space communications using both optics and radio, offering ultra-broad spectral resources 0. **Free-space optics (FSO)** has recently emerged as a potential candidate to complement radio technologies for both terrestrial and space communications, including fixed mobile x-haul links (front-, mid-, and backhaul), ground-to-satellite communications, and inter-satellite communications. Conventional FSO communications in the near-infrared (NIR) telecom band, though technologically mature for deployment by reusing fiber-optic telecom components and devices, face several challenges that impact their performance and practicality. These include atmospheric attenuation, weather sensitivity, such as susceptibility to particle scattering and turbulence effects, which can degrade signal quality, and eye safety concerns due to the potential harm caused by NIR wavelengths. To date, these issues have hindered the wide deployment and scalability of FSO communications to be a reliable part of the ICT infrastructure. Consequently, the research community explores and proposes other spectral regions for FSO applications.

Among various spectral regions, the mid-wave infrared (MWIR, 3-5 μm, 60-100 THz) and long-wave infrared (LWIR, 8-12 μm, 25-37 THz) in the mid-infrared (MIR) region hold significant potential for FSO technologies. They offer lower atmospheric propagation attenuation, broader unlicensed bandwidth, higher resilience against adverse weather conditions, and lower eye safety risks compared to other frequency bands 0. However, the MWIR and LWIR spectral regions for FSO applications have been underexploited due to the lack of efficient and effective transceiver technologies, as the semiconductor lasers and detectors operating at the NIR telecom band cannot be straightforwardly reused. Currently, the research community observes encouraging efforts and results in the research and development of MWIR and LWIR FSO communications from both the device and the system level. High data rate transmissions are made possible in these spectral regions with various approaches, progressing at an even faster pace than fiber-optic communications in recent years.

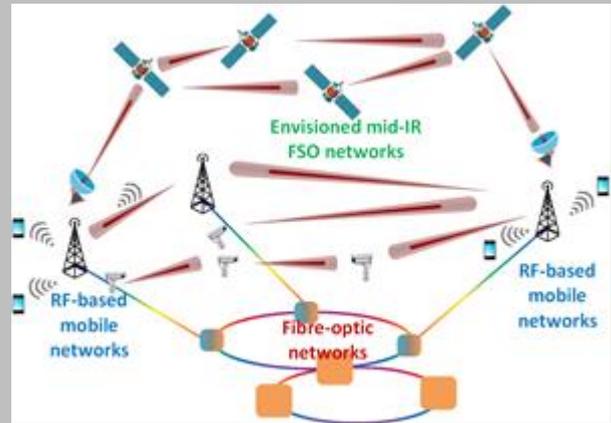

**Fig. 1.** Envisioned mid-IR FSO networks as a part of next-generation ICT infrastructure operating in parallel with the fibre-optic and the RF-based mobile networks.

## II. State of the art

There are two primary photonic approaches to generating and detecting signals for free-space communications in the MWIR/LWIR range: wavelength conversion and directly emitting laser sources.

The wavelength conversion-based mid-IR FSO systems utilize well-developed telecom transceivers and difference frequency generation (DFG) to convert signal wavelengths between 1.5 μm and mid-IR. While these systems have reported high data rates, they suffer from high-power consumption, energy deficiency, and hardware complexity, hindering their practical development. The wavelength conversion approach reuses mature optical and optoelectronic components from fiber-optic systems for free-space transmissions. Wavelength conversion from the telecom band is achieved using nonlinear parametric conversions in periodically poled $LiNbO_3$ (PPLN) devices. This allows the transmitter and receiver to operate directly in the telecom band, e.g., 1.55 μm. With this approach, data rates of MWIR free-space links can keep up with fiber-optic systems, achieving up to 300 Gb/s aggregated data rates 0. Advantages include very high data rates, compatibility with fiber-optic systems, and quick installation for high-end applications. However, drawbacks include limitations to the sub-4 μm region, high pump power requirements, low energy efficiency, and bulky PPLN devices.

The second approach, which involves direct emission laser sources and modulators in the MWIR and LWIR regions, provides numerous benefits such as superior energy efficiency, easy integration, and the absence of power loss due to frequency conversion. One early example of this technology utilized a PbCdS diode laser operating at a wavelength of 3.5 μm to transmit data at a rate of 100 Mb/s. Direct-emission semiconductor components and devices, including quantum cascade lasers (QCLs), interband cascade lasers (ICLs), external Stark-effect modulators, quantum cascade detectors (QCDs), and quantum well IR photodetectors (QWIPs), among others, present more promising long-term solutions by enabling compact semiconductor transceivers for free-space communications. QCLs operate using inter-subband transitions, which cover a broad wavelength range extending from the MIR to terahertz (THz) regions. These lasers exhibit significant advancements in terms of wide bandwidth, high-temperature operation, and low energy consumption.

**Table I.** Summary of lab demonstrations of quantum/interband cascade devices-enabled free-space transmissions

| Wavelength (μm) | Data rate/ bandwidth [a] | Year |
|---|---|---|
| 7.3 | 10 MHz [b] | 2001 |
| 9.3 | 330 MHz / 115 kb/s | 2001 |
| 8.1 | 2.5 Gb/s | 2002 |
| 8.1 | 750 MHZ-1.5 GHz | 2002 |
| 73 | 580 kHz | 2007 |
| 10.46 | 20 kHz | 2008 |
| 3 | 70 Mb/s | 2010 |
| 72.6 | 1 Mb/s | 2013 |
| 77 | 5 Mb/s | 2013 |
| 92 | 20 Mb/s | 2015 |
| 4.7 | 40 MHz | 2015 |
| Mid-IR [c] | 20 MHz | 2015 |
| 10.6 | 1 Gb/s | 2019 |
| 4.65 | 3 Gb/s | 2017 |
| 4.65 | 4 Gb/s | 2017 |
| 4.65 | 6 Gb/s | 2021 |
| 4 | 680 Mb/s | 2022 |
| 9.6 | 11 Gb/s | 2022 |
| 8.6 | 12 Gb/s | 2022 |
| 4.1 | 16 Gb/s | 2022 |
| 9.15 | 8.1 Gb/s | 2022 |
| 9 | 30 Gb/s | 2022 |

[a] The data rate for digital transmission, and modulation bandwidth for analog transmission are listed, respectively; [b] modulated on to a 66 MHz carrier; [c] wavelength not explicitly specified.

Directly modulated (DM) QCLs are particularly noteworthy due to their ultra-short carrier relaxation lifetimes, which result in high intrinsic modulation bandwidth and an over-damped laser response. This suppression of resonance frequency makes DM QCLs suitable for a wide array of applications, including free-space communications. Since the early 2000s, several demonstrations based on DM QCLs have reported data rates reaching up to a few Gb/s. After the invention of THz QCLs in 2007, transmission experiments achieved data rates of up to tens of Mb/s also in the THz region. While these initial demonstrations required cryogenic temperatures, subsequent developments enabled room-temperature broadband modulation of QCLs in the MWIR and LWIR regions. Consequently, numerous QCL-based free-space transmission demonstrations at room temperature have been reported.

Furthermore, recent efforts involving alternative schemes, such as directly modulated ICLs and external Stark-effect modulators, have yielded encouraging and promising results. These advances in direct emission laser sources and modulators for the MWIR and LWIR regions continue to pave the way for more efficient, high-performance, and versatile free-space communication systems.

Table I summarizes representative lab demonstrations with solid-state direct-emission MIR transceivers, including directly modulated QCLs 0, directly modulated ICLs, and external Stark-effect modulators 0. It is evident that there is a growing momentum in the field of direct emission transceivers for free-space communication systems, with rapid advancements being made in recent years. Based on the current trend, it is reasonable to anticipate that data rates exceeding 100 Gb/s will be achieved with direct emission transceivers in the near future. This progress will help bridge the gap between MIR FSO systems and fiber-optic systems. These recent developments present a promising trajectory for the evolution of FSO technology, addressing the ever-increasing demand for high data rates in next-generation wireless communication technologies such as 6G. By offering a more energy-efficient and practical solution, direct emission transceivers in the MWIR and LWIR regions will play a crucial role in shaping the future landscape of wireless communication systems. As the technology continues to mature, further research and development will be necessary to optimize the performance of direct emission transceivers, ensuring their ability to meet the complex and diverse requirements of various applications. Nonetheless, the ongoing progress in this area instills confidence that these emerging technologies will be instrumental in shaping the future of high-speed, energy-efficient, and robust free-space optical communication networks.

### III. CHALLENGES AND FUTURE WORKS

#### A. High-efficient transceiver technologies for system integration and miniaturization

The development of efficient, high-performance, and reliable transmission, modulation, and detection devices in the MWIR and LWIR bands to match the maturity level of fiber-optic NIR transceivers remains a significant challenge. Specifically, there is a need for effective phase modulation and coherent detection in these bands to enable the use of complex modulation formats in digital coherent optical transmissions. This necessitates the invention of novel devices with advanced capabilities.

One such capability is wavelength tunability, which allows the transceiver to operate over a range of wavelengths,

enhancing the flexibility and adaptability of the system. This is crucial for optimizing spectrum utilization with wavelength division multiplexing (WDM), accommodating various transmission distances, and potentially supporting dynamic FSO network reconfiguration. Developing tunable devices in the MWIR and LWIR bands requires innovative approaches, such as leveraging new materials or implementing unique device designs, to overcome the inherent limitations of existing technologies. Additionally, high-temperature operation capability is essential for the deployment of MWIR and LWIR systems in demanding environments, such as industrial settings, aerospace, or defense applications. Devices with high-temperature operation capabilities can withstand harsh conditions and maintain optimal performance, thereby improving the overall reliability and robustness of the communication system. Achieving this capability necessitates research into temperature-resistant materials, improved thermal management techniques, and the development of device architectures that minimize the impact of temperature variations on performance.

Therefore, to advance MWIR and LWIR technology towards the maturity level of fiber-optic NIR transceivers, a multidisciplinary approach will be required, combining expertise in material science, device engineering, and optical communication systems, to ensure that the developed technologies meet the stringent requirements of next-generation FSO communication networks.

B. *Reinvent and develop active/passive optical components for MWIR and LWIR*

In addition to the transceivers, the technology readiness level (TRL) for other essential devices operating in the MWIR and LWIR bands is relatively lower compared to those in other spectral windows. Developing mature components and systems for these bands demands substantial research and investment in various aspects of optical communication technology.

Many active and passive optical components that are crucial for setting up viable FSO networks, such as amplifiers, filters, (de-)multiplexers for both wavelengths and polarizations, and wavelength-selective switches, need to be reinvented for MWIR and LWIR operation. This entails overcoming a range of technical challenges, such as designing new materials and device structures that are compatible with these spectral bands, ensuring adequate performance, and maintaining cost-effectiveness. Amplifiers, for instance, must be redesigned to provide sufficient gain and low noise levels in the MWIR and LWIR bands. This may involve investigating novel gain materials or developing new pumping schemes to optimize amplifier performance in these bands. Filters and (de-)multiplexers must be adapted to function effectively in the MWIR and LWIR bands, providing precise wavelength selectivity and low insertion loss. This may require exploring new filtering mechanisms or leveraging advanced fabrication techniques to create compact, high-performance devices.

Wavelength-selective switches (WSS), which are vital for flexible and dynamic FSO network routing, must also be developed for the MWIR and LWIR bands. Achieving this goal may involve researching new switching technologies or optimizing existing designs to ensure efficient, fast, and reliable operation in these spectral regions. Moreover, efforts to increase the TRL of MWIR and LWIR devices should encompass improving their durability, power efficiency, and ease of integration into existing and future optical communication systems. This will necessitate close collaboration between researchers, engineers, and industry stakeholders to ensure a smooth transition from laboratory prototypes to commercially viable products.

C. *Optimal modulation, coding and multiplexing schemes for MWIR and LWIR*

The MWIR and LWIR bands are situated between the RF and NIR fiber-optic wavelengths, which results in an optimal trade-off between bandwidth and signal-to-noise ratio (SNR). This balance affects various aspects of communication, such as serial versus parallel transmission, single-carrier versus multi-carrier systems, and binary versus multilevel formats.

In addition to the optimal modulation and coding schemes as determined by information theory, other practical considerations must be taken into account when configuring modulation, coding, and multiplexing for MWIR and LWIR FSO systems. Factors such as resilience against adverse weather conditions, for example, will play a crucial role in the design and implementation of these systems. Particularly, atmospheric turbulence can still impact signal quality and overall performance of MWIR and LWIR FSO systems. To mitigate these effects, cooperative transmission of FSO with RF in the millimeter-wave (MMW) or THz band can be considered. This approach allows for the combination of the benefits of both FSO and RF technologies, providing more robust communication links that can adapt to different atmospheric conditions and maintain high data rates. Moreover, adaptive techniques can be employed to optimize the communication system's performance under varying conditions. These may include dynamic adjustments to the modulation and coding schemes or the implementation of advanced error correction techniques to improve the system's resilience against atmospheric turbulence and other environmental challenges.

D. *Free-space routing for NLOS scenarios*

Lastly, a wide range of applications calls for support in non-line-of-sight (NLOS) use cases. NLOS situations can lead to various challenges, such as signal obstruction, absorption, scattering, and multipath effects, which can significantly impact the performance and reliability of communication systems. To address these challenges, researchers are exploring several potential avenues, including:

- *Beam steering and tracking*: Implementing advanced techniques to precisely control the direction and angle of the transmitted beams, enabling them to bypass obstacles and maintain strong communication links despite NLOS conditions.

- *Adaptive optics for MWIR and LWIR wavelengths*: Developing new optical technologies that can adapt to changing environmental conditions and mitigate the effects of atmospheric turbulence, scattering, and absorption, ensuring robust and reliable communication links in NLOS scenarios.
- *Multi-hop and/or relay-assisted communications*: Utilizing multiple intermediate nodes or relay stations to relay signals between the transmitter and receiver, circumventing obstructions and maintaining connectivity in NLOS situations.
- *Distributed multiple-input multiple-output (D-MIMO) systems with joint transmission functions*: Implementing MIMO technology, which employs multiple transmitters and receivers located separately to send and receive multiple data streams simultaneously, in combination with joint transmission functions to coordinate and optimize data transmission in NLOS environments.

While these potential solutions offer promising avenues to tackle the challenges associated with NLOS use cases in MWIR and LWIR FSO systems, further research and development are necessary to refine these strategies, enhance their effectiveness, and ultimately improve the overall performance and reliability of NLOS FSO communications. Such research will help ensure that MWIR and LWIR FSO systems can effectively support a wide range of applications in diverse and challenging environments.

### REFERENCES

[1] X. You *et al*, "Towards 6G wireless communication networks: vision, enabling technologies, and new paradigm shifts," *Science China Information Sciences*, vol. 64, no. 1, pp. 110301:110301–110301:110374, 2020.

[2] A. Delga and L. Leviandier, "Free-space optical communications with quantum cascade lasers," *Proc. Quantum Sensing and Nano Electronics and Photonics XVI*, 2019, pp. 1092617.

[3] K. Zou *et al.* "High-capacity free-space optical communications using wavelength- and mode-division-multiplexing in the mid-infrared region," *Nat Commun*, vol. 13, no. 1, 2022.

[4] M. Joharifar *et al.*, "High-speed 9.6-µm long-wave infrared free-space transmission with a directly-modulated QCL and a fully-passive QCD," *J. Lightwave Technol.*, vol. 41, no. 4, pp. 1087-1094, 2023.

[5] H. Dely *et al.*, "10 Gbit s−1 free space data transmission at 9 µm wavelength with unipolar quantum optoelectronics," *Laser & Photonics Reviews*, pp. 2100414, 2021.

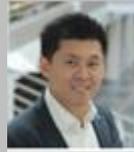
**Dr. Xiaodan Pan**g is a Docent at KTH Royal Institute of Technology, Sweden. He received his Ph.D. degree from Technical University of Denmark in 2013. He worked as a Post Doc at RISE Research Institutes of Sweden and as a researcher at KTH ONLab. From 2018 to 2020, Dr. Pang was a Staff Opto Engineer at the Infinera HW R&D as the PI of the EU H2020 MSCA-IF NEWMAN Project. He returned to KTH in 2020 upon receiving a Swedish Research Council (VR) Starting Grant. His research focuses on high-speed transmission technologies in MMW/THz, FSO and fiber-optics. He has authored over 200 publications and has given over 20 invited talks. He has been a TPC member of over 20 conferences, including OFC, OECC, ACP, and CLEO-PR, and he was the S1 subcommittee chair for OFC 2023. He is a Senior Member of IEEE and OPTICA and a Board Member of IEEE Photonics Society Sweden Chapter.

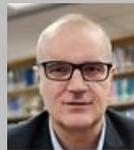
**Prof. Carlo Sirtori** received the Ph.D. degree in physics from the University of Milan in 1990. Following his degree in 1990, he joined Bell Labs where he started his research career on quantum devices. At Bell Labs he made important contributions in the field of semiconductor quantum structures such as the invention and the development of the "Quantum Cascade Laser". In 1997, he joined the THALES Research & Technology (TRT) in France. In 2000, he was appointed head of the "Semiconductor Laser Group" at THALES. Since 2002, he was Professor at the University Paris Diderot, and in 2010 he became Director of the MPQ laboratories of Paris Diderot. Since 2018 he is professor with Ecole normale supérieure and holds the ENS-THALES Chair of the Centre of Quantum Devices. He has received several prestigious awards such as the Fresnel Prize (European Physical Society) and various prizes in the USA, such as the "quantum devices award". In 2010, he was awarded an ERC-advanced-grant for his pioneering research on quantum devices.

# Review of Low-Earth Orbit Satellite Quantum Key Distribution

Davide Orsucci, Amita Shrestha and Florian Moll
*German Aerospace Center (DLR), {Davide.Orsucci; Amita.Shrestha; Florian.Moll}@dlr.de*

I. INTRODUCTION

The progress in quantum computation hardware is threatening information technology and communication infrastructure as we know it, since quantum algorithms could break the security of currently used public-key cryptographic standards. This threat should be addressed urgently since an adversary could collect massive amounts of encrypted data with the aim of decrypting it once that becomes possible, which is the so-called store-now decrypt-later attack. Two classes of solutions are prominently under consideration to overcome this threat: the use of post-quantum cryptography and **quantum key distribution (QKD)**. Here, we address the QKD approach to solving this problem, since it has the advantage of guaranteeing information-theoretic security and providing long-term secrecy of the exchanged messages.

The implementation of QKD protocols requires optical and photonic building blocks, including dedicated transmitters and receivers that are capable of generating and measuring the quantum states. These elements are currently rather costly, even though their technology readiness level has steadily increased over the past years and several commercial-off the shelf solutions are available today. The most prominent obstacle to the widespread deployment of QKD is the fact that quantum communication can be performed over optical fibres for a distance that are at most a few hundred kilometres, even under the most ideal experimental conditions. This is a consequence of the fact that quantum signals cannot be amplified; therefore, their intensity decays exponentially along an optical fibre and is quickly overcome by noise.

Satellite-based QKD (Sat-QKD) is poised to play a fundamental role in pushing forward the widespread use of this quantum communication technology, as it is currently the only practically viable method for connecting QKD users who are very far apart. Long-distance links can be established using **free-space optical (FSO)** communications between satellites and optical terminals on Earth. Furthermore, many system building blocks can be inherited from classical space laser communication systems where technology readiness level is already considerably high. Presently, most of the satellite QKD designs focus on prepare-and-measure (PM) protocols with satellites in low-Earth orbit (LEO), therefore we here focus on this scenario. Other choices of protocols are possible, such as entanglement based or measurement-device-independent methods, but these are significantly more challenging in terms of needed technology. The oldest and most widely investigated PM QKD protocol is the so-called BB84 protocol and it is likely to be the favoured candidate to be employed in future Sat-QKD missions and network constellations [1].

II. STATE OF THE ART

A. Trusted node LEO satellite QKD

Near-term Sat-QKD constellations are expected to be used in a trusted-node configuration. In this setting, each satellite is employed to create secure, quantum-generated keys with the authorised user on the ground when a direct line-of-sight with the specified user is available. Suppose that two specific users, customarily called Alice and Bob, already have established a secure key with a satellite and that, at a later moment, they wish to establish a common secure key between them. This can be done by letting the satellite broadcast (e.g., via radio-frequency communication) a value corresponding to the XOR of Alice's and Bob's key, as is illustrated in **Error! Reference source not found.**. This can be interpreted as using Alice's key to encrypt via one-time-pad Bob's key and securely forwarding it to Alice. Alice can then recover Bob's key by XOR-ing the message with her own private key. Note that this architecture relies on the trustworthiness of the satellite node to maintain the overall security of the key exchange.

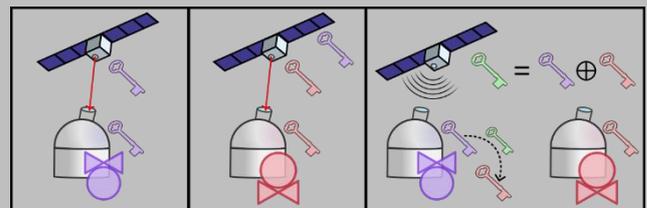

**Fig. 1.** Concept of satellite-based quantum key distribution to two ground nodes.

B. Review of recent satellite QKD activities

In the last few years, the interest in Sat-QKD has increased considerably, with many research institutions around the globe pushing to realise and launch LEO satellites that can enable QKD downlinks. Some notable activities include the following ones.

The feasibility of space links was investigated early on at the Matera Laser Ranging Observatory, by demonstrated single-photon level sensitivity by exploiting corner retro-reflectors mounted on a satellite [2]. First experiments of quantum communications from a mobile platform were conducted with a quantum transmitter installed in an aircraft and the receiver installed in DLR's Optical Ground

Station Oberpfaffenhofen [3]. Another early satellite quantum communications experiment was performed using the SOCRATES (Space Optical Communications Research Advanced Technology Satellite) satellite from NICT (Japan), whereby a QKD-like transmitter system was tested for space-to-ground communication [4]. The first full-fledged Sat-QKD demonstration was performed by the Chinese Academy of Sciences employing the Micius satellite. It successfully demonstrated QKD downlink with different ground stations, allowing the exchange of cryptographic keys between Asia and Europe. Furthermore, it demonstrated entanglement based QKD over a distance of 1200 km [5]. The European Space Agency (ESA) has also commissioned studies to assess the potential of Sat-QKD, leading to the kick-off of the SAGA project and of the EAGLE-1 satellite demonstrator. Furthermore, several national aerospace agencies are undertaking the construction of Sat-QKD demonstrators, including for instance QUBE (Germany), QEYSSat (Canada), QT Hub mission (UK), SpeQtral (Singapore), and several more [1].

### III. CHALLENGES AND FUTURE WORKS

The realisation of Sat-QKD terminals must comply with the stringent size, weight, and power (SWaP) requirements of satellite platforms. A promising path to reducing SWaP is the use of photonic integrated chips, which can integrate a complex set of passive and active optical elements (including lasers, phase and amplitude modulators, filters, and more). A fundamental element is the laser terminal that contains the antenna and the beam steering systems. Especially important for the Sat-QKD system is a high gain on the space side to close the link budget and at the same time keep the ground station receive aperture size within tolerable dimensions.

Existing optical ground stations (OGS) will need to be modified to host QKD receiver modules. Contained detectors need to have high sensitivity at the wavelength of choice. The detectors may be single-photon avalanche diodes or superconducting-nanowire single-photon detectors; the latter showcase very high detection efficiencies (often exceeding 90 %) but are still complex and bulky systems requiring cryogenic cooling. In the case of single-mode fiber interfaces maturity and efficiency of adaptive optics systems need to be increased. The main contribution to the transmission loss stems from the beam divergence since (even for diffraction-limited beams) a satellite-to-ground link results in a spot on the ground that is much larger than the aperture of existing OGS; other factors include atmospheric scattering and absorption, free-space-coupling or fibre-coupling of the incoming photons, pointing inaccuracy, and internal receiver and detector inefficiencies. Thus, link losses will realistically be at least -40dB. Eventually this also triggers the need of high rate QKD transmitters to generate sufficient key material.

The operating wavelength Sat-QKD should be selected to optimise the performance and minimise the system complexity. Typical choices exploit the atmospheric transparency windows either in the near-infrared band (NIR, around 850nm) or in the C-band (around 1550 nm). The NIR band suffers from increased background light and slightly higher atmospheric absorption, but it has less diffraction losses; furthermore, silicon detectors operating in the NIR band tend to have lower noise levels than the InGaAs detectors required for operations in the C-band. The C-band main advantage is the fact that it can benefit from the availability of many commercial off-the-shelf solutions developed for optical fibre classical communication.

REFERENCES

[1] J. S. Sidhu, S. K. Joshi, M. Gündoğan, T. Brougham, D. Lowndes, L. Mazzarella, et al., "Advances in space quantum communications," *IET Quantum Communication,* vol. 2(a), pp. 182-217, 2021.

[2] P. Villoresi, T. Jennewein, F. Tamburini, M. Aspelmeyer, C. Bonato, R. Ursin, C. Pernechele, V. Luceri, G. Bianco, A. Zeilinger and C. Barbieri, "Experimental verification of the feasibility of a quantum channel between space and Earth," *New J. Phys*. 10, 2008.

[3] S. Nauerth, F. Moll, M. Rau, C. Fuchs, J. Horwath, S. Frick and H. Weinfurter, "Air-to-ground quantum communication," *Nature Photonics,* vol 7, pp. 382–386, 2013.

[4] A. Carrasco-Casado, H. Kunimori, H. Takenaka, T. Kubo-Oka, M. Akioka, T. Fuse, Y. Koyama, D. Kolev, Y. Munemasa, and M. Toyoshima, "LEO-to-ground polarization measurements aiming for space QKD using Small Optical TrAnsponder (SOTA)," *Optics Express*, vol. 24, pp. 12254-12266, 2016.

[5] Y. Juan, Y-H. Li, S-K. Liao, M. Yang, Y. Cao, L. Zhang, J-G. Ren, et al., "Entanglement-based secure quantum cryptography over 1120 kilometres," *Nature*, vol. 582, pp. 501-505, 2020.

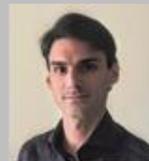

**Dr. Davide Orsucci** studied theoretical physics at the University of Pisa, Italy, attending in parallel to the program of the Scuola Normale Superiore, also located in Pisa (2008-2014), graduating with a thesis on topics at the intersection of quantum computation and cryptography. Afterwards, he did a PhD in quantum computation under the supervision of Prof. Hans Briegel in Innsbruck, Austria (2015-2018), investigating measurement-based quantum computation, quantum algorithms and quantum metrology. Subsequently, he did a post-doc in Basel, Switzerland, where he worked in the group of Prof. Nicolas Sangouard (2019) on topics related to quantum optics. He has since then joined the German Aerospace Centre (DLR) in the Quantum Communication Group led by Florian Moll, working on quantum key distribution, free-space optical communication, atmospheric channel modelling and quantum repeater architectures (2020-present).

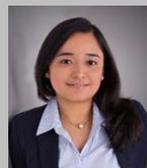

**Amita Shrestha** from German Aerospace Center (DLR), received her bachelor's degree in Electronics Engineering and communications from Kathmandu University, Nepal in 2006. She received her masters' degree in Communications Systems and Electronics form Jacobs University, Bremen in 2009. Since 2010 she is working in the Institute of Communications and Navigation in DLR focusing on Free space Optical Communications. In DLR, she has been involved in the development of real-time tracking software of Institute's optical ground stations, and its operation during several satellite and aircraft downlink experiments like OPALS, SOTA, DoDfast, VABENE etc. Currently, she is actively involved in the standardization of optical links in CCSDS (The Consultative Committee for Space Data Systems) community and several other projects related to free space optical classical and quantum communications. Additionally, she is chairing one of the working groups in COST Action CA19111

NEWFOCUS (European Network on Future Generation Optical Wireless Communication Technologies, 2020-2024) focusing on long range free space optical communication links.

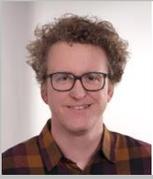

**Florian Moll** received his Dipl. Ing in electrical engineering from the Jena University of Applied Sciences in 2006 and a M.Sc. degree in electrical engineering from the Technische Universität München (TUM) in 2009. He has been a member of the German Aerospace Centre (DLR) Institute of Communications and Navigation since then. His work area is free-space optical quantum communications and telecommunications for aircraft and satellites. His main research interests are the connections between LEO and GEO satellites, aircraft and ground stations, characterization of the propagation channel and optics design. He was and is involved in several research projects as project leader and team member, dealing with classical and quantum communications. Since 2020, he is head of the research group Quantum Communications Systems

# Perspectives for Global-scale Quantum Key Distribution via Uplink to Geostationary Satellites

Davide Orsucci, Amita Shrestha and Florian Moll

*German Aerospace Center (DLR), {Davide.Orsucci; Amita.Shrestha; Florian.Moll}@dlr.de*

## I. Introduction

In this white paper, we analyse an unconventional satellite quantum key distribution (Sat-QKD) design where a single large satellite in geostationary orbit (GEO) allows the distribution of secure keys to many users. This design involves the use of many parallel uplink quantum channels, each supporting a prepare-and-measure **quantum key distribution (QKD)** protocol, such as BB84. Hundreds of users in the field-of-view (FOV) of the GEO satellite may be served simultaneously and using a wide FOV telescope may provide coverage to a whole continent at each time. By adjusting the satellite's pointing the field-of-regard can be further extended and all users located within the Earth's hemisphere visible from the GEO satellite could eventually establish a QKD link with it. Therefore, a single satellite could ultimately provide keys to a vast number of users, potentially hundreds of thousands, a task that would otherwise necessitate a large constellation of tens or hundreds of low-Earth orbit (LEO) satellites.

The goal of this approach to Sat-QKD is to shift expenses from the users to the satellite QKD provider and obtain cost savings from parallel access from many users. However, given the very large distance between a GEO satellite and the ground (35,790 km - 41,590 km depending on the specific location of a user on Earth), a very large aperture telescope, similar in size to the Hubble space telescope, is required to close the link budget, i.e., to obtain a non-zero secure key rate generation. Such telescope systems are currently extremely expensive, in the billion Euros range. However, the advantage of using such a telescope is that would be able, in principle, to serve many users (hundreds or thousands) simultaneously, while in most QKD concepts with LEO satellites each satellite can be linked to only one user on the ground at a time.

## II. State of the Art

The approach to Sat-QKD hereby presented is novel and deviates from conventional designs. Thus, there is no existing literature on this topic and this section will therefore focus on a preliminary concepts and ideas, which may lead to future more in-depth investigations.

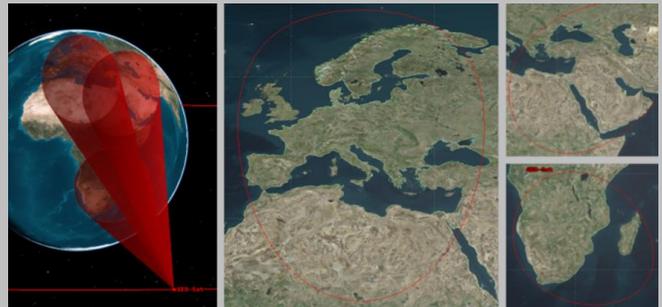

**Fig. 2.** Illustration of the field of view of the telescope, which can cover a continent at a time. By periodically adjusting the pointing of the telescope, the field of regard can be extended to cover almost half of Earth's surface.

### A. Advantages of employing a geostationary orbit

Using an uplink to a GEO satellite for QKD offers several advantages. User only need to own laser terminals, which are cheaper than single photon detectors. Also, since the GEO satellite maintains a nearly constant position in the sky, only fine-steering mirrors are needed for pointing the terminal; costs are reduced, compared to QKD with LEO satellites, since a non-moving modestly sized laser terminal (25 cm aperture) suffices.

Another advantage is that links to GEO satellites have a small point-ahead angle (around 20 µrad), which is typically within the atmospheric isoplanatic angle. This allows for effective compensation of beam wander in the uplink by tracking the apparent position of the downlink beacon sent from the satellite. As a final advantage, GEO satellites may have a long lifetime due to the absence of atmospheric drag. The lifetime limit could be determined by mirror degradation, or by the evaporation of coolant if evaporative cooling is required for maintaining the detectors at cryogenic temperatures.

### B. Satellite design

The concept design requires that the telescope is fit with several single-mode fibre (SMF) couplers located in the focal plane. The optical fibres then lead to a single-photon detecting system hosting hundreds of channels (i.e., detectors) in parallel. Each set of four channels would be used to implement an independent receiver front-end for a polarisation-encoded BB84 link. The satellite would be used in a trusted-node configuration, similarly to the case presented in the chapter "Review of low-Earth orbit satellite quantum key distribution". To streamline the presentation, we only consider 1550nm as QKD wavelength, in which the atmosphere is very transparent and for which mature laser communication technologies already exist.

To achieve positive QKD generation rate, the telescope aperture diameter needs to be very large (around 2 m). The telescope configuration could be, e.g., an off-axis three-mirror anastigmat (TMA); the off-axis TMA design effectively corrects all major aberrations and avoids having a central obscuration. As a result, the point-spread function (PSF) is ideally described by a Bessel function, leading to higher SMF coupling efficiency. In contrast, a central obscuration would cause the PSF to have an irregular shape, resulting in lower coupling efficiency. We consider a 3.1° semi-aperture viewing angle, which corresponds to approximately 8 square degrees of FOV.

The hundreds of individual beams sent in uplink by the users are convoyed towards optical fibre couplers located into the focal plane. An active switch and routing system, based for instance on micro-electromechanical mirror system or liquid variable lenses, needs to be employed to steer each individual beam to a fibre coupler.

After coupling to SMFs, the optical signals are conveyed to the single photon detectors required for the quantum communication. We suggest using superconducting nanowire single-photon detectors (SNSPDs). Though complex, these detectors offer high detection efficiency and low dark counts. At present, there are only a few theoretical articles addressing the use of SNSPDs in space [1] and achieving the required cryogenic temperatures in space is challenging. However, provided that a cryostat is available on-board, it should be relatively straightforward to accommodate hundreds of SNSPD channels within it.

*C. Link budget estimation*

We present a link budget estimation for the considered scenario, including the most relevant factors that influence the communication quality between the ground station and the satellite. The results are summarised in Table 1. The transmitter (Tx) antenna gain is calculated by assuming a truncated Gaussian beam with a waist (beam radius at $1/e^2$ of the peak intensity) of 15 cm in a terminal with a 25 cm diameter aperture, using the approach from [2]. The channel loss is determined for the challenging case where the user sees the satellite at an elevation angle is 30° and the link distance is around 38000 km. For nominal atmospheric visibility conditions (23 km) the atmospheric absorption is rather small. For nominal turbulence strength (Hufnager-Valley 5/7 turbulence model) the loss due to beam spread is substantial, even if the beam wander is fully corrected [3]. The receiver (Rx) antenna gain is determined assuming a telescope having a 2 m clear aperture. Several mirrors will be required for steering the beam, incurring in some optical losses even if highly reflective gold-coated mirrors are employed. Fibre coupling efficiencies exceeding 67% can probably be achieved [4], since the laser light wavefronts in uplink are almost perfectly planar.

TABLE 1. Link Budget estimation for the considered scenario.

| Parameter | Value [a] | Reference |
|---|---|---|
| Wavelength | 1550 nm | assumption |
| Tx antenna gain | 112.3 dB | Ø 25cm, waist 15cm, M2: 1.2 [2] |
| Free space loss | -289.8 dB | distance 38600km |
| Atmospheric loss | -0.7 dB | el. 30°, visibility 23km |
| Beam spread loss | -3.4 dB | el. 30°, HV 5/7 [3] |
| Rx antenna gain | 132.2 dB | Ø 2.0m |
| Rx transmission loss | -0.5 dB | 8 gold-coated mirrors |
| Rx coupling loss | -1.7 dB | 67% efficiency [4] |
| Reference transmission | -51.3 dB | table I in [5] |
| Total transmission | -51.7 dB | computed |

We avoid here giving specific details about realistic QKD transmitters and receivers and simply refer to the experimental work presented in [5]. For this QKD system it has been shown that at a link loss of -51.3 dB it is possible to generation of secure key rate block of 8.2 Mbit in a reasonably short time (1.17 hours). The key size is selected so that finite size effects only marginally reduce of the secure key generation rate.

III. CHALLENGES AND FUTURE WORKS

We have established that with the presented system design around -50dB of source-to-sink loss could be achieved for GEO uplinks. This is compatible with allowing the distribution of a few Mbits of key material to each connected user in around 1 hour. Therefore, this design constitutes a potential blueprint for achieving global-scale QKD with a single satellite. However, further investigations are needed to confirm the technological feasibility of the design and to consolidate the predicted system performance. Furthermore, the very significant costs entailed by placing a Hubble-class telescope in GEO make this proposal potentially realisable only in a rather long-term perspective.

REFERENCES

[1] P. Hu, Y. Ma, H. Li, Z. Liu, H. Yu, J. Quan, Y. Xiao, et al., "Superconducting single-photon detector with a system efficiency of 93% operated in a 2.4K space-application-compatible cryocooler," *Superconductor Science and Technology*, vol. 34, no. 7, 2021.
[2] B. J. Klein and J. J. Degnan. "Optical antenna gain. 1: Transmitting antennas," *Applied optics*, vol. 13, no. 9, 2134–2141, 1974.

[3] D. Federico, J. Antonio Rubio, A. Rodríguez, and A. Comeron, "Scintillation and beam-wander analysis in an optical ground station-satellite uplink," *Applied optics,* vol. 43, no. 19, pp. 3866-3873, 2004.

[4] N. Jovanovic, C. Schwab, O. Guyon, J. Lozi, N. Cvetojevic, Frantz Martinache, S. Leon-Saval, et al. Efficient injection from large telescopes into single-mode fibres: Enabling the era of ultra-precision astronomy. *Astronomy & Astrophysics,* vol. 604, A122, 2017.

[5] A. Boaron, G. Boso, D. Rusca, C. Vulliez, C. Autebert, M. Caloz, M. Perrenoud, G. Gras, et al. "Secure quantum key distribution over 421.4km of optical fiber," *Physical review letters*, vol. 121, no. 19, pp. 190502, 2018.

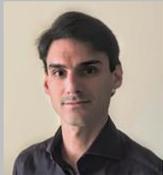
**Davide Orsucci** studied theoretical physics at the University of Pisa, Italy, attending in parallel to the program of the Scuola Normale Superiore, also located in Pisa (2008-2014), graduating with a thesis on topics at the intersection of quantum computation and cryptography. Afterwards, he did a PhD in quantum computation under the supervision of Prof. Hans Briegel in Innsbruck, Austria (2015-2018), investigating measurement-based quantum computation, quantum algorithms and quantum metrology. Subsequently, he did a post-doc in Basel, Switzerland, where he worked in the group of Prof. Nicolas Sangouard (2019) on topics related to quantum optics. He has since then joined the German Aerospace Centre (DLR) in the Quantum Communication Group led by Florian Moll, working on quantum key distribution, free-space optical communication, atmospheric channel modelling and quantum repeater architectures (2020-present).

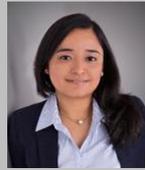
Aerospace Center (DLR), received her bachelor's degree in Electronics Engineering and communications from Kathmandu University, Nepal in 2006. She received her masters' degree in Communications Systems and Electronics form Jacobs University, Bremen in 2009. Since 2010 she is working in the Institute of Communications and Navigation in DLR focusing on Free space Optical Communications. In DLR, she has been involved in the development of real-time tracking software of Institute's optical ground stations, and its operation during several satellite and aircraft downlink experiments like OPALS, SOTA, DoDfast, VABENE etc. Currently, she is actively involved in the standardization of optical links in CCSDS (The Consultative Committee for Space Data Systems) community and several other projects related to free space optical classical and quantum communications. Additionally, she is chairing one of the working groups in COST Action CA19111 NEWFOCUS (European Network on Future Generation Optical Wireless Communication Technologies, 2020-2024) focusing on long range free space optical communication links.

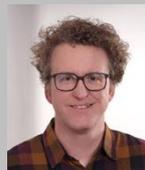
**Florian Moll** received his Dipl. Ing in electrical engineering from the Jena University of Applied Sciences in 2006 and a M.Sc. degree in electrical engineering from the Technische Universität München (TUM) in 2009. He has been a member of the German Aerospace Centre (DLR) Institute of Communications and Navigation since then. His work area is free-space optical quantum communications and telecommunications for aircraft and satellites. His main research interests are the connections between LEO and GEO satellites, aircraft and ground stations, characterization of the propagation channel and optics design. He was and is involved in several research projects as project leader and team member, dealing with classical and quantum communications. Since 2020, he is head of the research group Quantum Communications Systems.

# Wavelength Division Multiplexing Free Space Optical Links

Giulio Cossu, Veronica Spirito, Michail P. Ninos and Ernesto Ciaramella
*Scuola Superiore Sant'Anna, TeCIP Institute, Italy*
{giulio.cossu, veronica.spirito, michail.ninos, ernesto.ciaramella}@santannapisa.it

## I. Introduction

Nowadays **free space optical communications (FSOC)** are entering into the domain of satellite communications thanks to their shorter wavelength, providing much higher capacity than the traditional radio frequency (RF) wireless communications. High-speed satellite links are becoming of paramount importance for different purposes such as Earth observation, disaster recovery, last mile access of white areas, and they should exploit FSOCs to establish different types of links, such as feeder links (FLs), inter-satellite links (ISLs) or deep space links. This scenario is presented schematically in Fig. 1.

The use of optical beams shows other several benefits, such as the large modulation bandwidth, the high directivity, the reduction of the required space, weight, and power (SWaP) compared to RF-based systems. Last, but not least, optical beams can prevent interference between near carriers by using the unlicensed spectrum of the optical frequencies.

Leveraging upon the huge technological developments in fiber communications, the step forward can be accomplished with non-negligible, yet limited, technological effort with practical results of cost/benefit analysis. Very high-throughput traffic can be achieved by means of the well-known wavelength division multiplexing (WDM) technique. This is schematically represented in Fig. 2, where all WDM channels are combined and amplified by a booster erbium-doped fiber amplifier (EDFA) and launched into free-space by the telescope. After free-space propagation, the beam is collected by the receiver telescope, coupled into a single-mode fiber and amplified by a low-noise EDFA pre-amplifier. Finally, the channels are demultiplexed and the signals are detected by the corresponding optical receiver architecture. Using this type of links, FSOC with Terabit/s capacities were demonstrated terrestrial links [1]; they can now be taken to the satellite communications.

To reach this target, however, relevant innovations still must be introduced, such as new telescopes, interfacing free-space and fiber-based transceivers, and, of course, space-qualification of existing photonic components. However, the most challenging issues are related to the channel in FLs, where light is passing through the atmospheric layers, either in uplink or downlink. The atmosphere can attenuate the intensity and distort the wavefront of the optical signals, on millisecond timescales, leading to unpredictable fading events. To counteract these impairments, adaptive optics systems along with optimized acquisition and tracking methods should be deployed at the optical ground station (OGS) site. Here, we provide an overview of the emerging topic of WDM-FSOC in satellite networks and the key elements that must be considered when designing the links. We also highlight the technical gaps that should be addressed.

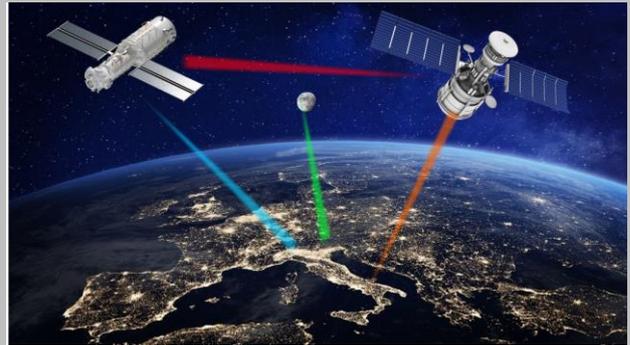

**Fig. 1.** Pictorial view of the different FSOC links: two FLs, one ISL and one deep-space link.

## II. State of the Art

Space agencies and private companies are heavily investing for developing prototypes to establish ISLs and the most challenging FLs. NASA project TeraByte InfraRed Delivery (TBIRD) aims in demonstrating a direct LEO-to-ground downlink (FL) laser communication link form a CubeSat towards a special ground terminal by NASA, providing a maximum data rate of 200 Gbit/s [2]. This high data rate is achieved by leveraging on off-the-shelf fiber-based coherent transceivers directly exploited in space applications. Recently, NASA announced that TBIRD completed the first batch of measurements, transmitting to ground 1.4 TB of data in a single pass of about five minutes (~ 40 Gb/s, on average).

The Lunar Laser Communications Demonstration (LLCD) was undertaken by NASA to demonstrate FSO between Moon and Earth. LLCD consisted of a duplex laser communication between a satellite in lunar orbit and a ground terminal on Earth, the Lunar Lasercom Ground Terminal (LLGT), a transportable system that was located in New Mexico. The communication was based on the optical C-band, the tests demonstrated the feasibility of a downlink transmission at 622 Mb/s and 20 Mb/s for the uplink. Sophisticated technologies were developed and used in these tests, such as superconductive nanowire detectors. The same ground station will support the Laser Communication Relay Demonstration (LCRD) [3], a long-term mission to demonstrate a bidirectional 1.2 Gb/s optical link between the ground and geosynchronous equatorial orbit (GEO) satellite.

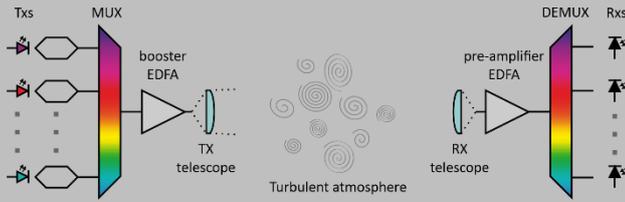

**Fig. 2.** Scheme of a typical transparent WDM-FSO system.

The European Space Agency's (ESA) "High thRoughput Optical Network" (HydRON) initiative aims to develop a "fiber in the sky" network providing seamless connectivity among different types of satellites, in various orbits, with the terrestrial networks: HydRON should integrate both space and ground segments into a high-capacity network [4]. This involves considering various payload configurations and network structures, while also considering different orbital characteristics, such as low Earth orbit (LEO) and GEO. Mission requirements, payload specifications, platform limitations, ground segment requirements, and network/protocol requirements should be taken into account, and preliminary designs are developed for the different scenarios under investigation.

Japan Aerospace eXploration Agency's (JAXA) Japanese Data Relay System (JDRS) program for Earth observation is already using an FSOC inter-orbit link between a spacecraft in low orbit and the data relay satellites. The new satellite carries a laser utilizing communication system by JAXA. It uses the infrared light to facilitate inter-satellite links at a data rate up to 1.8 Gb/s. The satellite will operate in a geostationary orbit around 36000 km altitude, relaying data between Japanese satellites. Data is downlinked via an RF Ka-band connection to the ground station.

In France, ONERA Labs are currently developing and integrating a novel OGS for FLs, addressing both GEO and LEO orbits, within the framework of the project FEELINGS. This OGS will be based on a 60 cm diameter telescope, embedding a high-power laser source, targeting 25 Gb/s DPSK links. The system should be launched by mid-2023 [5]. Another project by ONERA was the FEEDELIO experiment. Here, the aim was to demonstrate the effectiveness of the adaptive optics pre-compensation of atmospheric turbulence in presence of induced anisoplanatism. The demonstration has been performed from the Tenerife OGS and a terminal breadboard emulating a GEO satellite at a distance of 13 km, on top of mount Teide.

In Germany, other relevant experiments were conducted by the German Aerospace Center (DLR). As an example, the Optical Space Infrared Downlink System (OSIRIS) aims at developing downlinks from small satellites (and direct link between them). Depending on the configuration, the data rates range from 100 Mbit/s (for small satellite platform) to ~10 Gbit/s (for larger spacecrafts).

The Australian Space Agency (ASA) recently demonstrated in a field trial a potential solution to mitigate atmospheric effects, maintaining stable pointing to a moving target. They demonstrated a robust, high-speed coherent FSOC between an optical terminal and a drone moving with angular velocity comparable with the one of a LEO satellite. By integrating Machine Vision optical tracking and tip/tilt adaptive optics technology, both beam pointing and the angle-of-arrival were stabilized despite the presence of atmospheric turbulence. This allowed to maintain a horizontal link stable at 100 Gb/s.

### III. CHALLENGES AND FUTURE WORKS

In FSOC links, the limited divergence of the optical beam allows for higher irradiance at the RX but, at the same time, it increases the need for precise alignments, which must remain stable during the communication. This task is carried out by the pointing, acquisition, and tracking (PAT) subsystem. Nonetheless, the communication performance can be affected even under perfect alignment, due to the residual pointing jitter, which arises due to random angular fluctuations induced by the atmospheric turbulence and the unavoidable vibrations in mechanical structures.

In addition, in FLs, the optical signal passes through the atmosphere, suffering from several effects, shortly summarized in the following. Absorption and scattering produce a wavelength-dependent intensity attenuation that increases with decreasing the elevation angle and the OGS altitude. They are well-known deterministic effects, and their impact can be straightforwardly estimated.

The atmospheric turbulence is caused by wind and temperature gradients, which create eddies of air with varying densities and, thus, different refractive indices. These eddies can act like prisms and lenses, eventually producing a signal with an intensity that changes over time and space, an effect indicated as scintillation. To describe these intensity fluctuations, different statistical models have been proposed, but there is yet a lack of experimental evidence to strongly support any of them. Another important point to consider is the temporal nature of turbulence. In most practical cases, the channel fading changes very slowly, and the channel coherence time is usually between 0.1 and 10 ms.

Light passing through eddies on the order of (or larger than) the beam diameter (e.g., in feeder uplink), experience a phase modulation that combined with free-space propagation, results in a beam randomly moving around its boresight on the receiver plane, i.e., beam wander. Therefore, a beam emitted by an OGS towards a satellite can experience a significant angular displacement, on the order of several μrad. In downlink, the situation is different as the beam enters the atmosphere with a diameter much larger than the eddies of the atmosphere, because of the free-space beam expansion, and thus is affected minimally by beam wander effects.

In addition, in order to exploit the WDM technology in FSOC systems, the receiver telescopes must be able to couple the light into a single mode fiber. Under optimal conditions, the diameter of the focused beam should be the same as that of the single mode fiber guided mode, thus resulting in a coupling loss of 2-4 dB.

To (partially) counteract these effects, it is possible to act at communication or at hardware level. Alternative modulation/detection schemes can be exploited to increase the robustness, possibly combined with other common techniques (e.g., forward error correction). Moreover, the OGS should be equipped with adaptive optics, commonly used in optical astronomy. By employing this technique, the distortion induced in the signal wavefront by the turbulence is reduced by means of wavefront sensors and deformable mirrors. adaptive optics could allow to recover the signal wavefront in downlink and minimize the scintillation in uplink. In the last case, however, adaptive optics requires that the downlink and uplink signals have the same atmospheric path. Otherwise, anisoplanatic effects can limit the efficiency of adaptive optics subsystem. Unfortunately, importing adaptive optics imaging technology into the communications field is not straightforward since it requires much stricter requirements.

Finally, the background noise can also degrade the performance of the FSOC links. The receiver lens, especially in the terrestrial environment, collects also some undesirable and non-negligible background radiation, which can fall within the frequency range of the detectors as well as the optical bandpass filters, deteriorating the optical signal-to-noise ratio.

In conclusion, we presented an overview of WDM-FSOC links. The key open challenges are related to the actual implementation of such systems in real environments, since there is still lack of a reliable channel model, i.e., validated over different conditions. Crucial technological developments also include wide-size telescopes, with accurate PAT, powerful adaptive optics, and very good robustness against pointing errors. Another strategic element is the development of very high-power optical amplifiers, which can compensate for the losses, typically very high.

REFERENCES

[1] E. Ciaramella, et al., "1.28 Terabit/s (32×40 Gbit/s) WDM transmission system for free space optical communications," *IEEE J. Sel. Areas Commun.*, vol. 27, no. 9, pp. 1639–1645, Dec. 2009

[2] B. S. Robinson, et al., "TeraByte InfraRed Delivery (TBIRD): a demonstration of large-volume direct-to-Earth data transfer from low-Earth orbit," in *Free-Space Laser Communication and Atmospheric Propagation XXX*, 2018, vol. 10524, p. 105240V. doi: 10.1117/12.2295023.

[3] D. J. Israel, B. L. Edwards, R. L. Butler, J. D. Moores, S. Piazzolla, N. du Toit, and L. Braatz, "Early results from NASA's laser communications relay demonstration (LCRD) experiment program," Proc. SPIE 12413, Free-Space Laser Communications XXXV, 1241303 (15 March 2023)

[4] H. Hauschildt, et al., "HydRON: High thRoughput Optical Network," *Free-Space Laser Communications XXXII*, Mar. 2020, vol. 11272, p. 112720B. doi: 10.1117/12.2546106.

[5] P. Cyril, et al., "FEELINGS: the ONERA's optical ground station for Geo Feeder links demonstration," 2022 IEEE International Conference on Space Optical Systems and Applications (ICSOS), Kyoto City, Japan, 2022, pp. 255-260, doi: 10.1109/ICSOS53063.2022.9749705.

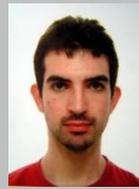
Giulio Cossu received M.S. degree in physics from University of Pisa (Italy), in 2010. He obtained his Ph.D. degree in 2014 at Scuola Superiore Sant'Anna (SA) of Pisa. Currently, he is Assistant Professor at SA. The main topic of his thesis was the investigation of innovative solutions of Optical Wireless Communications (OWC). His research interests include the areas of optical propagation through atmosphere, optical characterization, and optical communication. He was scientific responsible/technical officer for SA of the project "High Throughput Optical Network (HYDRON)" and "HYDRON Simulation TestBed", both founded by European Space Agency (ESA). He was in the workgroup about the development of optical wireless links for Intra/Extra Spacecraft and AIT scenarios within the framework of the TOWS project, founded by ESA. He is author or co-author of about 70 publications and holds 4 international patents.

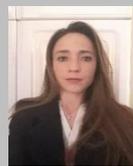
**Veronica Spirito** received the MS degree (cum laude) in Electronics Engineering from the La Sapienza University (Rome, Italy). She studied one year at the Universitat Politècnica de Catalunya in Barcelona, where she emulated a digital coherent communication system for FSO applications. Then, she moved to SA in Pisa. Here, as PhD candidate, she is involved in research activity applied to long-haul FSOC within several ESA research projects. As a system engineer, she is modelling the design of an end-to-end free-space WDM optical system for satellite communications. Focusing on the Physical Layer and signal propagation, she is currently at DLR German Aerospace Center (Munich, Germany) to conduct theoretical analysis and field trials of Earth-Satellites optical links.

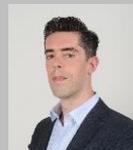
**Dr. Michail P. Ninos** received his Ph.D. in optical wireless communications from the National and Kapodistrian University of Athens, Greece, in 2019. In 2020 he joined IRIDA Research Centre for Communications Technologies at the University of Cyprus as a Postdoctoral Research Associate. From 2022, he is working as a Research Fellow at Scuola Superiore Sant'Anna, Pisa, Italy, under the ESA-funded project 'High Throughput Optical Network' (HydRON).

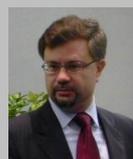
**Prof. Ernesto Ciaramella** is Professor of Telecommunications Scuola Superiore Sant'Anna Pisa, since 2002. His research interests include various areas of optical communications (components, systems, networks). His main research contributions are related to devices for the regeneration of the optical signal, the design of WDM systems for transport networks and access, and free-space optical systems. He is author or co-author of about 250 publications and holds 25 international patents. He participated in several European research projects. He was coordinator of the EU-FP7 project COCONUT (2012-2015), and is now principal investigator of ESA-TOWS project, about optical wireless systems.

# Review of Hybrid Optical-Radio Inter-Satellite Links in 6G NTN Including Quantum Security

Joan Bas and Marc Amay

*Centre Tecnològic de Telecomunicacions de Catalunya (CTTC), {joan.bas; marc.amay}@cttc.es*

## I. INTRODUCTION

Current mobile communication systems have started to integrate non-terrestrial networks to mobile systems (from Release 17 of 3$^{rd}$ generation partnership project (3GPP)). So, new applications such as direct-to-phone from a satellite are being intensively investigated in the framework of 3GPP and fifth generation (5G). In this scenario, mega-satellite constellations are being deployed /will be deployed to provide Internet-of-things (IoT) and broadband services according to the requirements of 5G/beyond5G/6G (sixth generation). These satellite constellations will provide global coverage, capacity and security as well as will complement the terrestrial infrastructure. These new satellite constellations will interconnect the satellites among them. The so-called inter-satellite-links (ISL). These links, according to 3GPP can be radio (millimetre waves) and optical, see [1].

Toward this regard, both SpaceX (Starlink), Amazon (Project Kuiper) and OneWeb, and Telsesat will use in their future mega-constellations of satellites optical inter-satellite links. On the contrary, Boeing plans to inter-connect the satellite of its constellation using mmWave links (V-band). Notice also that the satellite constellations equipped with inter-satellite links introduces a level of flexibility in the communications that permit to increase; the resilience to potential malfunction of the satellites; re-route the traffic to under-used satellites; extend the coverage with a lower number of satellites; avoid eavesdropping from tactical satellites; to develop multi-service constellations. Note that inter-satellite links may have different capacities. Thus, low, and high data rate services may be implemented in the same infrastructure (e.g., IoT, broadband connectivity, and others) and multiple operators can simultaneously operate in the same infrastructure. By doing so, the operating expenses, cost of maintenance, (operational expenditure (OPEX) and capital expenditure (CAPEX)) and the service cost are reduced.

## II. STATE-OF-THE-ART

The development of optical links to connect two satellites is quite recent. The first connection between two satellites using the optical bands was made in November 2001 between European Space Agency (ESA's) Artemis satellite located in geostationary satellite (GEO) orbit and Centre National D'Etudes Spatiales (CNES) Earth observation satellite called SPOT 4, which is in low Earth orbit (LEO) orbit [2] (a transmission distance of 40,000 km and a data rate of 50 Mbps). In 2005, ARTEMIS satellite began to be tested for bidirectional relay with the KIRAKI satellite of the Japanese space agency (JAXA) [3]. In 2014, ESA made the first link between satellites at gigabit speeds when the Alphasat TDP1 GEO satellites and the Sentinel-1A LEO satellites were connected [4]. Since then, multiple optical links have been made between the two satellites of a quasi-operational and experimental nature. The success of this system suggests that it may be applied to other links such as GEO-GEO (a transmission distance reaching up to 75000 km).

Similarly, the Sentinel 2A satellite was launched in October 2015 and was also equipped with inter-satellite optical links that were employed to connect with the Alphasat satellite. Investigations with the Sentinel 1A and 2A satellites gave way to what is known as the European data relay system (EDRS), which consists of a network of two GEO satellites that provide relay services to LEO satellites. The EDRS system has both optical inter-satellite and radio connection in the Ka band. The first ERDS satellite, called EDRS-A, is operated by Eutelsat (FR) and has been active since December 2016 and is known as EUTELSAT 9B EAST [5].

In November 2020 the Japanese agency JAXA carried out its own EDRS, which is called laser utilizing communication system (LUCAS), which made a link between a GEO and LEO satellites.

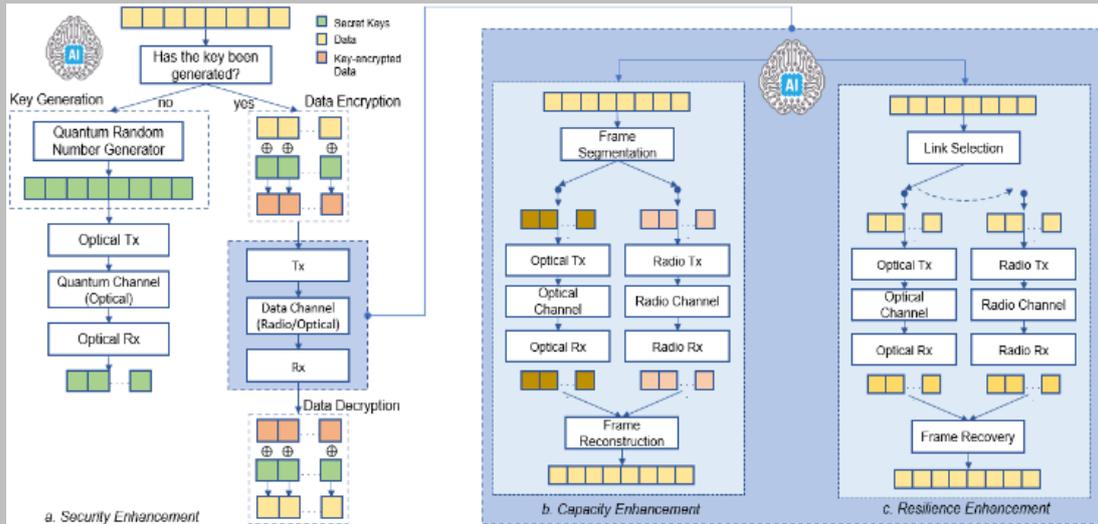

**Fig. 1.** Capacity, Resilient and Security schemes for hybrid optical-radio systems.

### III. CHALLENGES AND FUTURE WORK

Inter-satellite links allow communication between two satellites using either radio or optical frequency bands. However, its extension to satellite constellations presents several challenges including: (*i*) develop of low-cost coherent detectors; (*ii*) reduce the acquisition time of the links, especially for inter-orbit satellite links; (*iii*) to develop high-capacity systems but also resilient to impairments, see Fig. 1; and (*iv*) integrate the quantum links with the ISL links.

Toward this regard the future work may help to cope with the aforementioned challenges: (*i*) take advantage of the low temperatures of the space to integrate superconductivity elements to increase the capacity of radio and optical systems; (*ii*) introduce nanotechnology to miniaturize the radio and optical payload; (*iii*) introduce artificial intelligence to combine capacity and resilience communication schemes; (*iv*) develop quantum communication schemes for satellite with a large secrecy key rate by: decoupling the data and quantum channel using MEC techniques; resorting to signal processing combined with advanced coding schemes; use spatial diversity; utilize artificial intelligence to manage the quantum keys; and (*v*) extending the payload regeneration to include the quantum key distribution along a quantum satellite network.

### REFERENCES

[1] 3GPP TR 38.821 "3rd Generation Partnership Project; Technical Specification Group Radio Access Network; Solutions for NR to support non-terrestrial networks (NTN)", v16.0.0 (2019-12)

[2] A world first: Data transmission between European satellites using laser light" (http://www.esa.int/OurActivities/Telecomunications_Integrated_Applications/A_world_first_Data_transmission_between_European_satellites_using_laser_light). 22 November 2001.

[3] "Another world first for ARTEMIS: a laser link with an aircraft"(https://web.archive.org/web20090903020645/http://telecom.esa.int/telecom/www/object/index.cfm?fobjectid=27945). ESA. 19 December 2006.

[4] D. Tröndle; P. Martin Pimentel; C. Rochow; H. Zech; G. Muehlnikel; F. Heine; R. Meyer; S. Philipp-May; M. Lutzer; E. Benzi; P. Sivac; S. Mezzasoma; H. Hauschildt; M. Krassenburg; I. Shurmer, "Alphasat-Sentinel-1A optical inter-satellite links: run-up for the European data relay satellite system", In Proc. SPIE, Free-Space Laser Communication and Atmospheric Propagation XXVIII, vol 9739, pp. 973902, 15 March 2016; doi: 10.1117/12.2212744

[5] https://artes.esa.int/european-data-relay-satellite-system-edrs-overview , online; accessed 7/04/2023

[6] M.Amay, J. Bas, "On Hybrid Free-Space Optic-Radio Systems as Enablers of 6G Services over Non-Terrestrial Networks", In Proc. 2023 International Conference on Acoustics, Speech, and Signal Processing (ICASSP)

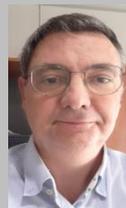

**Dr. Joan Bas** received his M.Sc and PhD (cum laude) in Electrical Engineering from the Universitat Politècnica de Catalunya (UPC). Currently, he holds at CTTC the position of Research Associate in the Space and Resilient Communication Systems (SRCOM) department. He has published +40 conference papers, journal papers, a book chapter, supervised Msc. and post-doc students, participated in +30 research projects, and +10 technological transfer ones. He is also serving as a reviewer of +15 journals. He has also participated in the organization of national and international conferences (e.g., ICASSP2020, GLOBECOM2021, Satellite Workshop on ICC 2022). His main research areas of interest are the improvement of the spectral and energy efficiency of satellite communications, indoor monitoring, optical wireless communications and its integration with RF systems, security on IoT systems and the design of integrated 6G Terrestrial and Non-Terrestrial Networks.

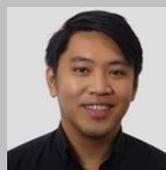

Marc Amay is a PhD Researcher at the Space and Resilient Communications and Systems (SRCOM) Research Unit of Centre Tecnològic de Telecomunicacions de Catalunya (CTTC) in Barcelona, Spain. He received his Bachelor degree in Electronics and Communications Engineering from New Era University, Philippines, then obtained his Double Master degree from Aston University, UK and Telecom Paris, France with distinction (cum laude). His research area of interest is on i.) Quantum Satellite Communications, ii.) Integrated Terrestrial-Non-Terrestrial Networks (3D Networks), and iii.) Hybrid Radio-Optical Wireless Communications, all to serve the paradigm of 6G communications.

# Highly-sensitive SPAD-based Optical Wireless Communication

Shenjie Huang[1], Mohammad-Ali Khalighi[2] and Majid Safari[1]

[1]The University of Edinburgh, UK, {Shenjie.huang; Majid.Safari}@ed.ac.uk

[2]Méditerranée, France, ali.khalighi@fresnel.fr

## I. INTRODUCTION

We have already witnessed the successful standardisation and deployment of the fifth generation (5G) wireless communication networks. However, the rapid emergence and evolution of smart products, interactive services, and intelligent applications result in a significant need for high-speed wireless communication. It is anticipated that the current 5G technology will be inadequate in meeting the demands of future communication requirements beyond the year 2030 and people have started to investigate what the future sixth generation (6G) will be. It is expected that 6G will provide key features including ultra-high speed, nearly 100% geographical coverage, sub-centimeter geo-location accuracy, enhanced energy efficiency, and high intelligence levels. These key features will greatly accelerate the development of the enabling technologies for smart cities such as the Internet of Things, autonomous vehicles, and intelligent transportation systems. All available spectra are expected to be explored in 6G to boost the data rates, including the sub-6 GHz, millimeter wave (mmWave), terahertz, and optical frequency bands. Therefore, **optical wireless communication (OWC)** is envisioned to play a pivotal role in future wireless networks.

Over the past few decades, the main applications of OWC, which have been extensively investigated within both academic and industry communities, include space communication, terrestrial free-space optical communication (FSO), indoor visible light communication (VLC), optical camera communication (OCC), and underwater OWC (UOWC). Despite having three orders of magnitude more spectrum resources compared to its radio frequency (RF) counterparts, OWC still faces significant challenges that must be addressed before it can be widely deployed, in particular, the occasional outages due to fluctuations in line-of-sight received power. Such power fluctuation can be introduced by various factors. Specifically, for indoor VLC, it can be caused by user mobility, device orientation, random blockage, and dimming control, while for outdoor FSO, it can result from turbulence effects, adverse weather conditions, and transceiver vibration. Highly sensitive receivers, such as **single-photon avalanche diodes (SPADs)** can be employed to effectively mitigate power outage issue and improve reliability. SPADs also have the important advantages of low cost, weight, and operation voltage, rendering them highly suitable for OWC applications.

## II. STATE OF THE ART

The photon counting capability of a SPAD is achieved by biasing the traditional linear photodiode (PD) above the breakdown voltage to operate in the Geiger mode. When a photon arrives, the SPAD can detect it and initiate an avalanche process which produces a striking current pulse. Compared to the traditional linear photodetectors such as positive-intrinsic-negative (PIN) PD and avalanche photodiode (APD), a SPAD has the advantage of significantly higher receiver gain (on the order of $10^6$). While SPAD-based receivers can offer single photon sensitivity, they need to be quenched after each avalanche triggered by a photon detection. During this quenching period, which is known as the *dead time*, the SPAD becomes temporarily insensitive to any incident photon arrivals, because the bias voltage of SPAD has not yet recovered to the operating level.

Depending on the adopted quenching strategies, SPADs can be categorised into two types, namely, passive quenching (PQ) and active quenching (AQ). AQ SPADs use active circuits to detect the avalanche pulse and then quench it, followed by resetting the SPAD after an accurate hold-off time, while PQ SPADs quench the avalanche currents by simply flowing through quenching resistors without using active circuitry. For PQ SPADs, any photon arrival during the dead time can extend its duration. In contrast, AQ SPADs exhibit constant and short dead times. Despite the superior performance of AQ SPADs, the current commercial and research and development (R&D) SPAD receivers still predominantly rely on PQ technology due to its cost-effectiveness and scalability.

SPAD has a wide range of applications such as in LiDAR, quantum communication, positron emission tomography and time-of-flight and fluorescence lifetime imaging. In recent years, there has been a growing interest in exploring the potential application of SPAD in the context of OWC. Specifically, in 2013, Fisher *et al.* introduced a reconfigurable 32×32 SPAD array receiver designed in standard 130 nm CMOS technology and studied its properties for VLC [1]. Later, numerous studies have been conducted to investigate SPAD-based VLC systems [2].

More recently, the applications of SPAD in UOWC and FSO were also studied in [3-4] and references therein. Although SPAD can greatly improve the receiver sensitivity in OWC and overcome outages caused by low received power, its performance is strongly limited by the dead time, especially in high received power scenarios. From the communication point of view, dead time mainly introduces two main impairments: nonlinear distortion [4] and intersymbol interference (ISI) [5].

Many research efforts have focused on mitigating the dead time effects and enhancing the throughput of SPAD OWC through the exploration of innovative receiver designs, decoding schemes, and signal-processing techniques. In particular, a hybrid SPAD/PD receiver which leverages both the high sensitivity of SPADs and the high data rate of linear PDs to achieve superior performance has been proposed in [4]. The hybrid receiver, as shown in Fig.1, consists of both PIN PD and SPAD array detectors and possesses dual operation modes. It can adaptively switch between the high sensitivity (SPAD) mode and the high data rate (PIN PD) mode depending on the incident optical power. Other works have explored the enhancement of performance improvement in SPAD OWC through signal-processing techniques, thus avoiding the need for additional hardware complexity. Because of the nonlinear nature of the dead-time-induced ISI, the commonly used linear equalization techniques cannot provide decent performance in SPAD OWC. Therefore, the adoption of nonlinear equalizers becomes crucial in high-speed SPAD OWC. The current record experimental data rate achieved by SPAD OWC in the open literature is **5 Gbps**, which was recently reported in [5]. The achievement of this high data rate was enabled by the integration of several techniques, including optical orthogonal frequency division multiplexing (OFDM), nonlinear equalization, peak-to-average power ratio optimization, and adaptive bit and energy loading.

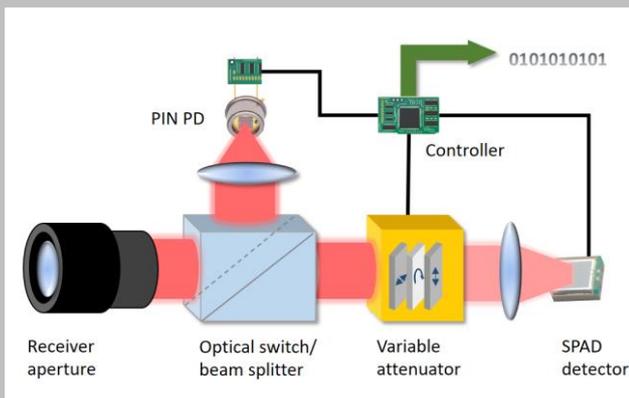

**Fig. 1.** The schematic diagram of the hybrid SPAD/PD receiver.

## III. CHALLENGES AND FUTURE WORKS

Although recent theoretical and experimental studies have already illustrated the feasibility and advantages of incorporating SPADs in OWC, further efforts are still needed to address the existing challenges before its commercialization and widespread adoption. SPAD is an imperfect photon-counting receiver due to the presence of some non-ideal effects. One critical challenge that needs to be overcome is the development of cost-effective SPAD detectors with superior specifications, such as short dead time, high photon detection efficiency, as well as low dark count rate, after-pulsing, and crosstalk probability. All these factors can affect communication performance and need to be improved to realize a high receiver sensitivity approaching the quantum limit. Especially, considerable research efforts should be devoted to developing highly efficient SPAD detectors operating in the infrared range, e.g., InGaAs/InP SPADs for 1550 nm, which are promising for long-distance OWC. In addition, the current research mainly focuses on the application of SPAD in VLC, FSO and UOWC. There is a notable absence of studies investigating its use in other OWC applications, such as OCC and optical vehicular communication. Therefore, it is imperative to conduct further research to explore the potential of SPAD in such applications.

Developing simple yet efficient techniques to combat the imperfections of SPADs, such as dead time and afterpulsing, is also a challenge that requires further investigation. Although several techniques have already been proposed in the literature, they suffer from some limitations, such as high complexity and limited efficacy. To address these issues, one promising research direction could be developing machine learning techniques tailored to SPAD-based OWC.

Most of the SPAD OWC works in the literature employed on-off keying as the modulation scheme, mainly because of the limited dynamic range of the SPAD receivers caused by dead time and small array size. Recently, because larger array SPAD detectors with low dead time become available, people have started to explore the application of higher-order modulation schemes, such as optical OFDM, in SPAD OWC. However, novel spectrum-efficient modulation schemes designed based on the unique properties of SPAD are highly desirable and are expected to achieve superior performance compared to traditional schemes. Other open challenges include the derivation of accurate photon counting models and the development of advanced decoding techniques. It can be anticipated that more rapid deployment of SPAD-based OWC is feasible if all the aforementioned challenges can be successfully addressed.

## REFERENCES

[1] E. Fisher, I. Underwood, and R. Henderson. "A reconfigurable single-photon-counting integrating receiver for optical communications." *IEEE Journal of solid-state circuits*, vol. 48, no. 7, pp. 1638-1650, 2013.

[2]. W. Matthews, Z. Ahmed, W. Ali, S. Collins. "A 3.45 Gigabits/s SiPM-based OOK VLC receiver." *IEEE Photonics Technology Letters,* vol. 33, no. 10, pp. 487-490, 2021.

[3]. M.A. Khalighi, H. Akhouayri, and S. Hranilovic. "Silicon-photomultiplier-based underwater wireless optical communication using pulse-amplitude modulation." *IEEE Journal of Oceanic Engineering*, vol. 45, no. 4, pp. 1611-1621, 2019.

[4]. S. Huang and M. Safari. "Reliable Optical Receiver for Highly Dynamic Wireless Channels: An Experimental Demonstration." 2021 IEEE Global Communications Conference (GLOBECOM), pp. 1-6, 2021.

[5] S. Huang, C. Chen, R. Bian, H. Haas, and M. Safari. "5 Gbps optical wireless communication using commercial SPAD array receivers." *Optics Letters*, vol. 47, no. 9, pp. 2294-2297, 2022.

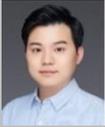
**Dr. Shenjie Huang** received the B.Sc. degree in optoelectronic engineering from Jiangnan University, China, in 2013, the M.Sc. degree in signal processing and communications from The University of Edinburgh, U.K., in 2014, and the Ph.D. degree in electrical engineering from The University of Edinburgh, U.K., in 2018. He is currently a Research Associate with the Institute for Digital Communications, The University of Edinburgh. His main research interest is optical wireless communications.

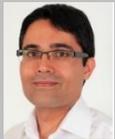
**Dr. Mohammad-Ali Khalighi** (Senior Member, IEEE) is Associate Professor with École Centrale Méditerranée, Marseille, France, and head of "Optical Communications for IoT" group at Fresnel Institute research lab. He is currently serving as Action Chair for the COST Action CA19111 NEWFOCUS (European Network on Future Generation Optical Wireless Communication Technologies). He was the Coordinator of the H2020 ITN MSCA VisIoN project (Visible-light-based Interoperability and Networking). He is also serving as Editor-at-Large for the IEEE Transactions on Communications, and has served as Associate Editor for the IET Electronics Letters as well as Lead Guest Editor for the IEEE Open Journal of the Communications Society. His main research interests include wireless communication systems with an emphasis on free-space, underwater, and visible-light optical communications.

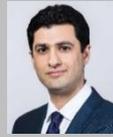
**Dr. Majid Safari** (S'08-M'11-SM'20) received his Ph.D. degree in Electrical and Computer Engineering from the University of Waterloo, Canada in 2011. He also received his B.Sc. degree in Electrical and Computer Engineering from the University of Tehran, Iran, in 2003, M.Sc. degree in Electrical Engineering from Sharif University of Technology, Iran, in 2005. He is currently a Reader in the Institute for Digital Communications at the University of Edinburgh. Before joining Edinburgh in 2013, He held postdoctoral fellowship at McMaster University, Canada. Dr. Safari is currently an associate editor of IEEE Transactions on Communications and was the TPC co-chair of the 4th International Workshop on Optical Wireless Communication in 2015. His main research interest is the application of information theory and signal processing in optical communications including fiber-optic communication, free-space optical communication, visible light communication, and quantum communication.

# On Signalling and Energy Efficiency of Visible Light Communication Systems

Tilahun Gutema and Wasiu Popoola

*School of Engineering, University of Edinburgh, UK. {tilahun.gutema; w.popoola}@ed.ac.uk*

## I. INTRODUCTION

A visible light communication (VLC) system that uses low cost light emitting diodes (LEDs) is a viable approach to complement the existing radio-based communication systems. Energy-efficiency, low-cost, and ubiquity of LEDs and a license-free spectrum make VLC a suitable candidate for low-power systems as well. However, achieving the potential of VLC requires a good understanding of the right signalling and waveform design for any given applications. In the case of resource-constrained wireless network, energy efficiency and robustness are paramount. A low-rate wireless network requires simple, low-cost, and reliable communication with limited power consumption. For such networks, energy efficiency is of significant importance to prolong battery life in applications such as internet-of-things (IoT). Equally vital is the robustness of the link to noise and other channel impairments.

Thus, a key challenge in designing and optimising a VLC system is to improve its energy and spectral efficiency [1]. Energy efficiency is important as it can help to reduce the power consumption and operating costs of VLC systems, while spectral efficiency is crucial for supporting many users or applications in a given frequency band. An important theoretical framework for understanding the relationship between signal-to-noise ratio (SNR), bandwidth, and the capacity of a communication channel is the ultimate capacity limit formulated by Shannon in 1948 [2]. According to Shannon's theory, the capacity of a communication channel is a linear function of the bandwidth, but a logarithmic function of the SNR. This means that increasing the bandwidth of a VLC system can significantly increase its capacity while increasing the SNR has a less pronounced effect. As a result, it is important to optimise not only the SNR but also the bandwidth of a VLC system to maximise its capacity.

## II. STATE OF THE ART

Significant research effort has been geared towards the development of high-speed VLC. Such systems have to address the challenges associated with the relatively low modulation bandwidth of commercial LEDs. Various techniques such as pre- and post-equalisation, high-order modulation signalling have been explored to optimise the spectral efficiency and achieve high data rates in the order of Gb/s using a single LED. It has also been demonstrated that the LED modulation bandwidth increases as bias current increases until it saturates. However, an LED is a nonlinear device. That is, its emitted optical power as a function of the driving current is not linear. Consequently, driving the LED in the nonlinear region distorts the transmitted signal and deteriorates the system performance. Therefore, it is imperative to optimise the DC bias point of an LED to benefit from increasing bandwidth at driving current while minimising any nonlinear distortion associated with operating beyond the dynamic range of the LED as reported in [3].

Figure 1 shows an example of this optimisation in terms of the achieved bit rate per channel use at different bias current for LED VLMB1500-GS08. This illustrates that as the bias current is increased, the increase in modulation bandwidth translates to increasing transmission rate but beyond the optimum point, the bandwidth increase is offset by the reduced SNR and increased nonlinearity/distortion.

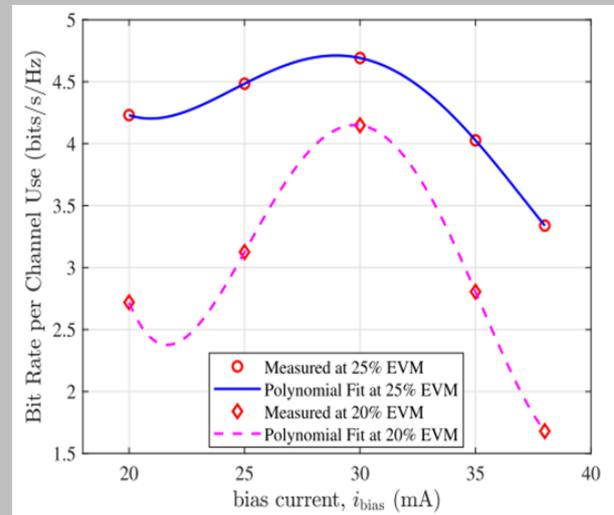

Fig. 1: Achieved bit rate per channel use considering 25% and 20% reference error vector magnitude (EVM) at different bias currents for LED VLMB1500-GS08.

Another approach is to optimise the SNR term and maximise the available channel capacity for any given bandwidth that is available for use. This can be realised by adjusting the distribution of source symbols using signal-shaping techniques. Conventional data transmission schemes transmit each symbol with equal probability, which is not optimal for the additive white Gaussian noise (AWGN) channel [4]. By using signal-shaping techniques, it is possible to optimise the distribution of source symbols and improve the spectral efficiency of a VLC system. This can help to close the gap between the spectral efficiency of a VLC system and Shannon's channel capacity. A viable way to achieving this is probabilistic shaping (PS).

The PS approach reduces the amount of symbol energy required in transmission. This is because PS provides a Gaussian-like distribution over the signal constellation. Therefore, low-energy symbols, which have a lower standard deviation in the Gaussian-like distribution, are transmitted more frequently than high-energy symbol, as graphically illustrated in Fig. 2. This reduction in average symbol energy at a specific error rate, compared to a uniform distribution, is known as shaping gain. The resulting shaping gain can be used to increase the SNR, which results in increased noise resilience and a higher achievable information rate. Thus, PS is a valuable technique for improving the performance of VLC systems, particularly in high-speed applications.

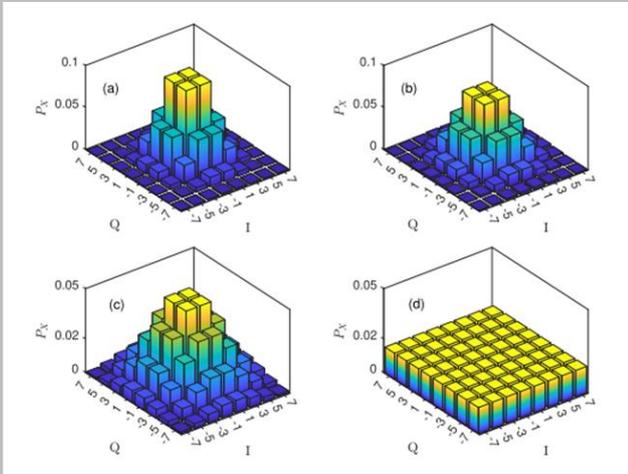

Fig. 2: Graphical illustration for probabilistic shaped with four different entropy values (a) H = 4.5 bit/symbol, (b) H = 4.80 bit/symbol, (c) H = 5.40 bit/symbol, (d) H = 6.00 bit/symbol (uniform distribution)

Signalling methods to achieving efficient use of available energy in VLC systems include orthogonal modulation schemes. These schemes have applications in low-power and low-rate networks. However, orthogonal modulations sacrifice spectral efficiency to approach Shannon's limit. As a result, it is important to carefully balance the trade-off between energy efficiency and spectral efficiency when designing VLC systems using orthogonal modulations.

Signalling techniques such as on-off keying (OOK), and $M$-ary pulse amplitude modulation (M-PAM) are simple to implement and have been studied extensively [5]. While it is possible to achieve a 1 bit/s/Hz spectral efficiency with OOK, it is generally regarded as energy inefficient. Spectral efficiency of $\log_2 M$ bit/s/Hz can be achieved with $M$-PAM but the required SNR increases with increasing $M$. On the other hand, in orthogonal modulation techniques such as $M$-ary pulse-position modulation ($M$-PPM) and frequency shift keying ($M$-FSK), the energy efficiency improves as $M$ increases at an expense of spectral efficiency. Moreover, PPM has a peak-to-mean optical power ratio and is very sensitive to receiver synchronisation that requires complex equalisation techniques to mitigate. On the other hand, FSK waveform has a constant envelope and requires relatively simpler equalisation methods. For these reasons, previous studies, such as [6], have considered the energy-efficient FSK based VLC for low data rate and low-power IoT applications. However, in frequency selective channels and applications where the modulation bandwidth is limited and the overall link end-to-end response is non-flat, such as in VLC, the performance of FSK-based systems begin to degrade. That is because FSK symbols mapped onto frequency region where the channel attenuation is larger will lead to erroneous detection.

### III. CHALLENGES AND FUTURE WORK

An emerging signalling approach for VLC is chirp modulation in which a symbol is mapped to all frequencies linearly within the bandwidth. Consequently, the chirp signalling is resilient as the symbol energy not concentrated in a single subcarrier signal frequency but spread across a progressively increasing range of frequencies within the symbol duration. As a result, the entire energy of a data symbol is somewhat shielded from severe attenuation suffered by any single subcarrier signal frequency. Thus, 'chirping' the symbol energy over a range of subcarrier signal frequencies offers resilience to the combined effects of the channel and limitation of the front-end devices.

The application of VLC in addressing the wireless communication challenges in the subsea channel continue to attract attention. Water absorption, multiple scattering and turbulence pose serious challenges for underwater VLC though. Other emerging trends in the VLC research include the use of machine learning in signal detection and channel mitigation and heterogeneous network that combines multiple bearers. However, for VLC to realise its potential, there will need to be a move away from the line-of-sight dominated studies to beyond line-of-sight VLC systems.

### REFERENCES

[1] A. Ozyurt, W. Popoola et al. 'Energy and spectral efficiency of multi-tier LiFi networks'. IEEE Wireless Comm. and Net. Conf. (WCNC), 2023.
[2] C. E. Shannon, "A mathematical theory of communication," *Bell System Technical Journal*, vol. 27, no. 3, pp. 379–423, 1948.
[3] T. Z. Gutema, H. Haas and W. O. Popoola, "Bias point optimisation in LiFi for capacity enhancement," *Journal of Lightwave Technology*, vol. 39, no. 15, pp. 5021-5027, 2021.
[4] F. Buchali, F. Steiner, G. B¨ocherer, L. Schmalen, P. Schulte, and W. Idler, "Rate adaptation and reach increase by probabilistically shaped 64-QAM: An experimental demonstration," *Journal of Lightwave Tech.*, vol. 34, pp. 1599–1609, 2016.
[5] Z. Ghassemlooy, W. Popoola, and S. Rajbhandari, Optical wireless communications: system and channel modelling with MATLAB®. CRC press, 2019.
[6] A. W. Azim, et al. 'Energy efficient M-ary FSK-based modulation techniques for VLC', *IEEE Transactions on Cognitive Comm. and Networking*, vol. 5, no. 4, 2019.

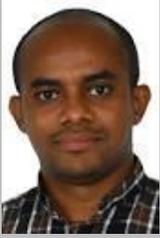
**Tilahun Gutema** received the B.Sc. degree in electrical engineering from Addis Ababa University, Addis Ababa, Ethiopia, in 2015, and the M.Sc. degree in optics and photonics from the Karlsruhe Institute of Technology, Karlsruhe, Germany, in 2018. He is currently working toward the Ph.D. degree with The University of Edinburgh, Edinburgh, U.K. His main research interests include digital modulation techniques and visible light communication.

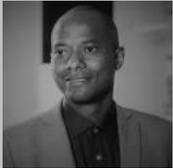
**Dr. Wasiu Popoola** is currently a Reader (Communication Engineering) in the School of Engineering, University of Edinburgh, U.K. He holds a prestigious RAEng/Leverhulme Trust Research Fellowship, he is a Fellow of the Institute of Engineering Technology and Higher Education Academy. He has authored or co-authored more than 120-journal articles/conference papers/patents, including the book Optical Wireless Communications: System and Channel Modelling with MATLAB (two editions). He is also a Science Communicator appearing in science festivals and on the 'BBC Radio 5live Science' programme in October 2017. He was an Invited speaker at various events including IEEE Photonics Society Summer Topicals

# Practical Implementation of Outdoor Optical Camera Communication Systems with Simultaneous Video and Data Acquisition

Vicente Matus, José Rabadán and Rafael Perez-Jimenez
*University of Las Palmas de Gran Canaria, Spain, {vicente.matus; jose.rabadan; rafael.perez}@ulpgc.es*

## I. INTRODUCTION

The utilization of visible light communication (VLC) systems, such as Li-Fi, which offer simultaneous illumination and data transmission using light-emitting diodes (LEDs), has primarily been confined to indoor applications where link spans are short and the interference caused by background lights is manageable. Outdoor implementations of VLC are not feasible due to the high intensity of sunlight and the need for longer link spans. However, optical camera communication (OCC) allows for intrinsic source and noise segregation using existing camera equipment, albeit at lower speeds [1].

In outdoor environments, OCC brings promise in supporting data reception for sensor networks, a use case that often requires low data rates, a large number of nodes, and a restricted energy budget. Furthermore, OCC systems can offer simultaneous video and data acquisition, a significant advantage in scenarios such as surveillance systems, as depicted in Fig. 1, where variables from a farming field, such as soil moisture, humidity, temperature, and others, can be monitored by sensors that transmit data using LEDs to a surveillance camera or a drone monitor. The dual data and video acquisition functionality is possible when certain conditions are met: the dimensions of the LED-based source at the transmitter side are small, the linkspan is long, and the camera field of view is wide. In these conditions, the LED's projection on the image sensor is of negligible area, leaving most of the image free for video monitoring.

Our study examines the novel sub-pixel OCC condition, which occurs when the transmitter size is less than the size of a pixel at the image plane. When the sub-pixel condition is satisfied, most of the information from the transmitter is captured by one single pixel, with adjacent pixels receiving an attenuated copy of the information due to camera de-focus or atmospheric scattering and turbulence.

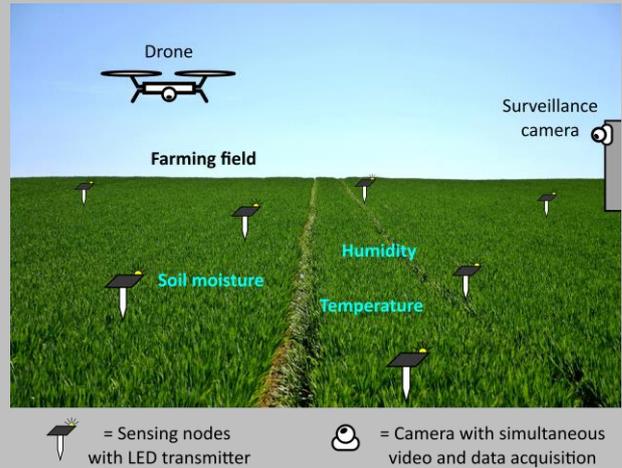

**Fig. 1.** Use case of optical camera communications allowing simultaneous video and data acquisition from light-emitting diode (LED)-enabled sensing nodes in a farming field.

## II. STATE OF THE ART

Since early implementations of OCC, most systems have exploited the effect of rolling-shutter (RS) acquisition in which the image sensor of the camera is exposed on a row-by-row basis [2], significantly improving the achievable data rate at the expense of the field of view and link span, leaving no room for the actual image or video capture during the communication process. In turn, global-shutter (GS) systems have been long neglected due to the scarcity of such type of image sensor, which has mainly been replaced by more cost-effective complementary metal-oxide semiconductor (CMOS)-based sensors.

The inherent capacity of cameras to separate multiple sources of light has been exploited to implement multiple-transmitter topologies, taking advantage of spatial division and multiple-input multiple-output (MIMO) techniques [4]. Using the camera as a multiple-photodiode receiver, it becomes possible to simultaneously transmit independent data streams from multiple transmitters, each associated with a different LED source. The camera can spatially divide and capture the individual signals, enabling parallel communication channels and increasing the system's throughput.

Outdoor OCC systems face challenging atmospheric conditions that degrade the link, such as turbulence, particles in the air, sunlight, and different types of precipitations. Nevertheless, thanks to the image-forming optics and the control over the photographic parameters of the camera, it has been demonstrated that such conditions can be efficiently overcome [5].

By combining the previously mentioned sub-pixel conditions, we have demonstrated that an on-off keying signal can be transmitted at distances in the range of one hundred meters. In Fig. 2, we show the signal obtained by a single pixel in these conditions. As can be seen in the image, the setup allows for simultaneous video and data acquisition, promising a practical way to implement sensor networks based on sub-pixel OCC.

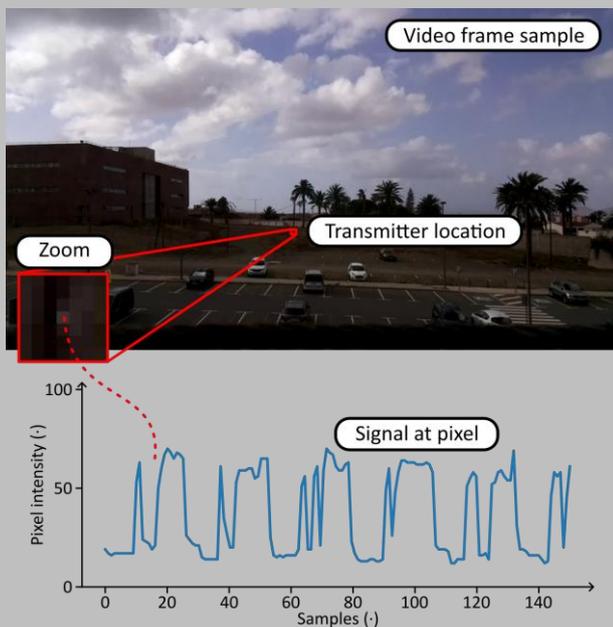

**Fig. 2.** Signal obtained from video footage in outdoor OCC with subpixel conditions

### III. CHALLENGES AND FUTURE WORKS

The main challenges in the implementation of sensor networks using OCC in outdoor environments are the energy budget for the transmitter and the discovery and tracking of nodes. LED transmitters can be energy efficient if oriented correctly with a narrow beam, but this limits the mobility of the nodes. On the other hand, mobile nodes would require high-power LEDs with isotropic radiation patterns, thus consuming more energy. Additionally, from the image processing perspective, the receiving nodes need to identify multiple nodes in the video stream and track them over time if they are moving. This requires adapting artificial intelligence-based computer vision models, trained for pattern recognition, to recognize and interpret the unique patterns created by the communication process.

Future works need to focus on developing efficient transmitters that can adjust the radiation pattern of the light source. The data rates and energy efficiency could also be enhanced by using M-ary pulse amplitude modulation schemes. At the receiver, the development of image processing algorithms needs to address pattern recognition, tracking, and image stabilization, combined with digital signal processing for the demodulation of the information. Cloud-based architectures should also be explored since the processing of heavy computer vision models can be delegated to a back-end instead of the computing module of the receiver. The protocols for video streaming and compression should be taken into account in the data detection process.

Furthermore, the extrapolation of subpixel OCC to short-range links holds the potential for supporting sensor networks in indoor facilities. The available LED indicators in most devices and machinery meet the subpixel condition for standard cameras at a few meters distance. This opens up new opportunities for applications such as indoor localization, smart home and office automation, and wearable devices, where subpixel OCC can provide efficient communication and monitoring capabilities.

### REFERENCES

[1] N. Saeed, S. Guo, K.-H. Park, T. Y. Al-Naffouri, and M.-S. Alouini, "Optical camera communications: Survey, use cases, challenges, and future trends," *Physical Communication*, vol. 37, pp. 100900, 2019.

[2] N. T. Le, M. A. Hossain, and Y. M. Jang, "A survey of design and implementation for optical camera communication," *Signal Processing: Image Communication*, vol. 53, pp. 95-109, 2017.

[3] Z. Ghassemlooy, P. Luo, and S. Zvanovec, "Optical camera communications," in Optical Wireless Communications: An Emerging Technology, Springer, 2016, pp. 547-568.

[4] S. Teli, S. Zvanovec, R. Perez-Jimenez, and Z. Ghassemlooy, "Spatial frequency-based angular behavior of a short-range flicker-free MIMO–OCC link," *Applied Optics*, vol. 59, pp. 10357-10368, 2020.

[5] V. Matus, V. Guerra, C. Jurado-Verdu, S. Zvanovec, and R. Perez-Jimenez, "Wireless sensor networks using sub-pixel optical camera communications: Advances in experimental channel evaluation," *Sensors*, vol. 21, no. 8, pp. 2739, 2021.

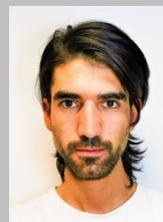

**Dr. Vicente Matus** received the degree of Electrical Engineer from the University of Chile in 2018, and his PhD from the University of Las Palmas de Gran Canaria (ULPGC) in Spain in 2021. He was a Marie S. Curie fellow in the Horizon 2020 program of the European Union for the ITN-VISION project. Currently, he is a researcher in the Photonics division of the Institute for Technological Development and Innovation in Communications (IDeTIC) at the ULPGC, and a visiting researcher at the Instituto de Telecomunicações Aveiro in Portugal, funded by the Catalina Ruiz 2022 scholarship from the Canary Islands Government. His research focuses on the experimental development and deployment of optical camera communication systems and their outdoor applications for wireless sensor networks. Among the applications of his interest are the automation of agriculture systems, vehicular communication, and wearable sensors for medical applications.

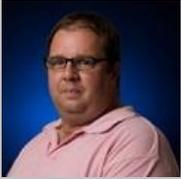

**Prof. José Rabadán**, M.S. in 1995 and PhD in 2000, both at the Universidad de Las Palmas de Gran Canaria, ULPGC, (Canary Islands, Spain). He is now a professor at the ULPGC and researcher in the IDeTIC (Institute for Technologic Development and Innovation in Communications). His research interests are in the field of wireless optical communications for both wideband local area networks and narrowband sensors networks, high performance modulation and codifications schemes for wireless optical communications. He has been also working on RF applications, mainly developing environmental intelligence networks and Internet of Things (IoT) applications based on WIFI and Bluetooth systems and RF identification devices (RFID). He has been a researcher in different national and international projects financed by local national and European Administrations and companies. He also is the author of several national and international technical publications (books, book chapters, papers in national and international journals and conferences).

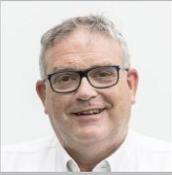

**Prof. Rafael Perez-Jimenez** received his M.Sc. degree in 1991 from Universidad Politécnica de Madrid, Spain, and his Ph.D. degree (Hons) in 1995 from Universidad de Las Palmas de Gran Canaria, Spain. He is a full professor at the ULPGC, where he leads the IDeTIC Research Institute. His current research interests are in the field of Optical Camera Communications (OCC), optical indoor channel characterization and the design of robust visible light communication (VLC) systems for indoor communications, especially applied for sensor interconnection and positioning. He has been awarded with the Gran Canaria Science Prize (2007), the Vodafone Foundation Research Award (2010) and the RSEAPGC Honor Medal (2017).

# Physical Layer Security for Visible Light Communications in the Era of 6G

Erdal Panayirci[1], Panagiotis D. Diamantoulakis[2] and Harald Haas[3]

[1]Department of Electrical and Electronics Engineering, Kadir Has University, Istanbul, Turkey,
eepanay@khas.edu.tr

[2]Department of Applied Informatics, University of Macedonia, 54636 Thessaloniki, Greece,
padiaman@ieee.org

[3]Department of Electronics and Electrical Engineering, The University of Strathclyde, Glasgow G1 1XQ, UK, harald.haas@strath.ac.uk

## I. INTRODUCTION

Security, privacy and trustworthiness by design are envisioned to be the key objectives of a novel class services in the era of the sixth generation (6G) of wireless networks [1]. A brief presentation of all the envisioned classes of services in the era of 6G is presented in Fig. 1. In order to achieve the key performance indicators (KPIs) of the classes of services in the era of 6G and beyond, the use of physical layer security (PLS), except of increasing the overall system's security, is expected to facilitate the provision of other classes of services in 6G, including ultra-massive machine-type communication, energy sustainable communication, and extremely reliable and low-latency communication. This is because the approach of physical layer security (PLS) can play a vital role in reducing both the latency as well as the complexity of novel security standards. In the meanwhile, it is expected that the dramatic increase in high data rate services will continue its trend to meet the demands of 6G networks. Thus, the use of PLS in communication systems that use higher frequency bands is of paramount importance, to mitigate the spectrum (especially of the convenient sub6 GHz frequency band) saturation, which also negatively affects PLS.

**Optical wireless communications (OWC)** and **visible light communications (VLC)**, offer attractive features such as high capacity, robustness to electromagnetic interference, a high degree of spatial confinement, inherent security, and unlicensed spectrum. Depending on the intended application, VLC can serve as a powerful complementary technology to the existing ones, such as the wireless body area network and personal area network, wireless local area network, vehicular area network, and underwater hybrid acoustic/VLC underwater sensor network and it will also be useful in scenarios in which traditional RF communication is less effective such as in airplanes, underwater communications, healthcare zones, etc.

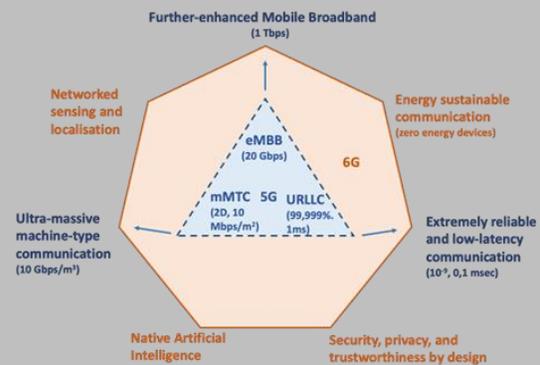

**Fig. 1.** 6G classes of services.

During the past few years, PLS in VLC networks has emerged as a promising approach to complement conventional encryption techniques and provide a first line of defense against eavesdropping attacks. The key idea behind PLS is to utilize the intrinsic properties of the VLC channel to realize enhanced physical layer (PHY) security, without reliance on upper-layer encryption techniques. We use the *secrecy capacity* as a performance measure to determine the maximum communication rate that guarantees reliable reception of the secret message by the authorized receiver.

The evolution toward 6G wireless communications poses new and technical challenges which remain unresolved for PLS in VLC research, including PLS coding, massive multiple-input multiple-output (MIMO), non-orthogonal multiple access, full duplex technology and so on. Moreover, it is not possible to employ conventional PLS techniques as in radio frequency (RF) communications [2]. In fact, the most practical communication scheme for VLC systems is intensity modulation and direct detection. Due to the nature of light, the intensity-modulating data signal must satisfy a positive-valued amplitude constraint. The state of the art, the challenges, and future works of the PLS enhancement methods for VLC are presented in this paper.

## II. STATE OF THE ART

PLS will play a vital role in enhancing cyber-security in wireless networks. Moreover, it will also help reduce both the latency and the complexity of novel security standards. The provision of user security is distributed across all layers of the open systems interconnect (OSI) model. The integrity and confidentiality of information is typically ensured by using secret and public key encryption methods. However, the strength of these techniques may be enhanced by reducing the attack surface. In this regard, the physical layer exposes significant vulnerabilities due to the broadcast nature of the wireless channel. PLS systems can leverage the challenges presented by wireless channels, such as fading and noise, to enhance secrecy. By utilizing the randomness inherent in the channel, PLS ensures that messages remain concealed from adversarial users. It is well known that if the eavesdropper is equipped with sufficient computational power, protocol security cannot guarantee the secure transmission of information.

PLS is based on a strong theoretical basis, which has been established at least five decades ago. In more detail, the mathematical framework for wiretap channels was introduced by Wyner in 1975, eliminating the need for a secret key to guarantee secrecy. Subsequently, in 1978, Csiszar and Korner demonstrated the existence of channel codes that can provide both robust transmission and secrecy simultaneously. Also, In 1978, Leung-Yan-Cheong et al. introduced the concept of secrecy capacity. The main advantage of using OWC technology for increasing PLS is that light does not propagate through opaque objects such as walls. Light beams are also very directional - think of a laser beam in the extreme case. Hence, light beams can be formed without the need for excessive signal processing efforts. Lenses and other optical components can be used to shape a beam. It is, therefore, possible to significantly reduce the possibilities of man-in-middle attacks in light fidelity LiFi compared to wireless fidelity (WiFi). Fundamentals and techniques of PLS, developed for RF channels involving wire-tap coding, multi-antenna, relay-cooperation, and physical layer authentication, cannot be applied directly to VLC channels. This is mainly because many standard specifications in transmission protocols and modulation schemes of VLC systems are quite different from RF systems. Besides, light can easily be confined spatially and, since there is no fading because the wavelength is significantly smaller than the size of the detector, the VLC channels become more deterministic.

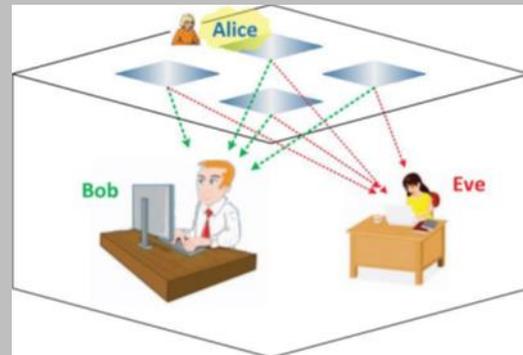

**Fig. 2.** An indoor SISO PLS scenario.

State of the art PLS techniques are mostly based on techniques such as jamming, mapping of transmitted symbols, precoding, and subset selection, as well as combinations of these techniques. A typical indoor single-input-single-output (SISO) PLS system is shown in Fig. 2. In addition, MIMO and wavelength division multiplexing (WDM) can be employed to enhance physical layer security. In this context, spatial modulation (SM) exhibits advantageous features due to its property to use the propagation channel for information transmission. In SM, the information is carried by the transmitted symbols, as well as by the indices of an active transmit unit. MIMO and MIMO-SM-based PLS systems were studied extensively in research and development work widely presented in the literature [3]. However, it is important to note that practical systems may be unaware of potential interceptions by passive eavesdroppers. Consequently, sub-optimal secure MIMO transmission is primarily achieved through transmitter preprocessing techniques such as precoding, friendly jamming, mapping, and selection.

In precoding approaches, widely adopted in most applications due to their simplicity, through the channel state information at the transmitter of the legitimate user, the precoding matrix coefficients are constructed through some optimization techniques so that the confidential message is perceived by the legitimate user clearly while the eavesdropper's bit error rate performance is degraded substantially. On the other hand, friendly jamming is based on the transmission of a signal that creates an artificial noise, which lies in the null space of the legitimate user. After combining the confidential information with the jamming signal at the transmitter side, only the eavesdropper will experience destructive effects from the jamming signal. In the secrecy enhancement techniques, the secrecy is realized by an encryption key for the given modulation. The same key is used on the legitimate user's side to decode the confidential message. Another PLS enhancement technique, called transmitter subset selection, is based on choosing a specific subset of transmitting entities according to the radiation patterns of the transmitting units.

The design of confidential signal sets is based on maximizing the minimum Euclidean distance or SNR at the legitimate user. Finally, the hybrid design of VLC and RF systems was expected to improve the user experience, substantially, since VLC systems can support reliable high data rates in specific areas and RF systems can provide coverage when a line-of-sight link is not available. They can co-exist, operating in the same environment, without causing any interference. it is also possible that both systems share the same physical layer techniques and medium access control (MAC) algorithms such as authentication and encryption. A transmitter luminaire communicates with the legitimate receivers in the presence of an external eavesdropper, which can be combined for hybrid RF/Optical PLS systems [4]. Since hybrid VLC/RF systems have both VLC and RF components altogether in the system, PLS for such systems should be jointly investigated due to the broadcast nature of both technologies. In more detail, to increase PLS in cross-band VL/RF systems, a joint resource allocation is needed, considering all the available resources and corresponding constraints, the overall CSI information and the users' quality-of-service requirements. For example, if a user requires a high level of PLS and is located near an LED AP, the system's resource management system may avoid using RF to serve this specific user.

### III. CHALLENGES AND FUTURE WORKS

VLC is a technology that will boost the emergence of next generation wireless networks. However, despite the inherent advantages of this technology regarding some specific features, e.g., security, compared to the RF counterpart technologies, the related weaknesses should be further investigated and mitigated. More specifically, the use of VLC does not imply the immunity of the communication system to security threats, due to the broadcast nature of VLC systems. Thus, further enhancing PLS in VLC systems can make VLC one of the core technologies for 6G applications that require a high level of cyber-security. Using some metrics such as the secrecy rates, the secrecy performance of VLC system can be significantly improved in various scenarios through well designed PLS techniques. A promising approach is based on the design of novel algorithms for PLS in multiuser and broadband VLC systems. To this direction, the use of new modulation schemes such as spatial modulation techniques, including index modulation and orthogonal frequency division multiplexing (OFDM)-index modulation, and optical multiple-input-multiple output techniques with non-orthogonal multiple is very promising. The algorithms to be designed must have high power efficiency and must have the capability to work in multi-user scenarios. In particular, the artificial jamming signal generation property of these modulation techniques is the most important advantage in providing PLS compared to the traditional approaches.

Also, the successful use of VLC to increase PLS in future generations of wireless networks depends on two issues: a) the successful collaboration of this technology with the prominent RF communication systems, and b) the potential to use visible light technology in a multi-fold way to simultaneously offer several functionalities, including secure communication, lighting, sensing and localization, and power transfer. Such an approach could have a major impact on several important real-world applications, including smart industry, distributed and cloud computing, extended reality, etc. In more detail, except the cross-band resource allocation that has been discussed in the previous section, another promising approach to increase PLS in cross-band VL/RF systems is by using VLC to increase the sensing and localization precision. This approach can directly improve security and trustworthiness since the physical location of each node is known. Also, the information acquired from sensing and localization can be used to optimize the beam steering in the RF-based MIMO subsystem. To this direction, the use of machine learning has shown important advantages compared to conventional optimization techniques. Another promising research direction is the investigation of PLS in cross-band systems with simultaneous lightwave and power transfer (SLIPT) [5]. For example, the jamming signals can be used both to increase the secrecy rate and for energy harvesting by internet-of-things devices, which in the uplink can report their data to the access point by using RF.

Consequently, the main research challenges for PLS in VLC are as follows:
- Which are the most suitable physical layer features to be exploited for the definition of security algorithms in 6G heterogeneous environments characterized by high network scalability and different forms of active malicious attacks?
- How can PLS and covert communication be provided in a proactive way and without requiring feedback from the wireless devices?
- How can PLS be provided for devices with extremely low energy consumption (e.g., IoT) or zero-energy devices and how can it be combined with other key technologies for the seamless operation of the aforementioned devices, e.g., simultaneous lightwave and power transfer?
- Which MAC protocols for VLC offer notable advantages in terms of PLS and which of them are suitable in different VLC applications?
- How much can the use of reconfigurable optical lenses (e.g., at the receiver) and intelligent surfaces improve PLS?
- How can artificial intelligence be exploited to tune physical layer security algorithms dynamically?
- How can lightweight key distribution and authorization techniques best be developed that leverage the previously obscured PHY-layer attributes while maintaining ultra-low latency (ULL) quality of service (QoS)?
- How can lightweight key distribution and authorization techniques best be developed that leverage the previously obscured PHY-layer attributes while maintaining ULL QoS?
- What relevant, unique dimension reduction/feature extraction methods could enable transfer learning while maintaining the privacy of aggregator networks over various RF interfaces?
- How can PLS be ensured with transmitters (i.e., LED access points) and photodiode-based receivers distributed in different locations?
- How can confidentiality be ensured between the central baseband processing unit and antenna stripes?
- How can user mobility and device orientation be incorporated into the VLC channel models and combining VLC and RF?

ACKNOWLEDGMENTS

The work of P. D. Diamantoulakis was supported by the Hellenic Foundation for Research and Innovation (H.F.R.I.) under the "3rd Call for H.F.R.I. Research Projects to support Post-Doctoral Researchers" (Project Number: 7280).

REFERENCES

[1] W. Tong and P. Zhu, Eds., *6G: The Next Horizon: From Connected People and Things to Connected Intelligence*, 1st ed. Cambridge University Press, 2021.
[2] M. A. Arfaoui, M. D. Soltani, I. Tavakkolnia, A. Ghrayeb, M. Safari, C. M. Assi and H. Haas, " Physical layer security for visible light communication systems: A survey", IEEE Communications Surveys & Tutorials, Vol. 22, No. 3, pp. 1887-1908, April 2020
[3] E. Panayirci, A. Yesilkaya, T. Cogalan H. Haas and H, V. Poor, "Physical-layer security with generalized space shift keying", *IEEE Trans. Commun*, vol. 68, no.5, pp. 3042 - 3056, May 2020.
[4] J. Liu, J. Wang, B. Zhang and Q. Wang, "Secrecy performance analysis of hybrid RF/VLC dual-hop relaying systems", *Optics and Photonics*, vol. 9, 2021
[5] G. Pan, P. D. Diamantoulakis, Z. Ma, Z. Ding and G. K. Karagiannidis, "Simultaneous Lightwave Information and Power Transfer: Policies, Techniques, and Future Directions," in *IEEE Access*, vol. 7, pp. 28250-28257, 2019.

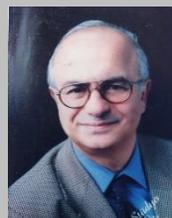

**Prof. Erdal Panayirci** (Life Fellow, IEEE) received a Ph.D. degree in electrical engineering and system science from Michigan State University, USA, in 1971. He is currently a Professor with the Department of Electrical and Electronics Engineering, Kadir Has University, Istanbul, Turkey, and a Visiting Research Collaborator with the Department of Electrical Engineering, Princeton University, Princeton, NJ, USA. He has published extensively in leading scientific journals and international conferences and coauthored the book Principles of Integrated Maritime Surveillance Systems (Kluwer Academic, 2000). His research interests include communication theory, synchronization, advanced signal processing techniques, and their applications to wireless electrical, underwater, and optical communications. He is a member of the National Academy of Sciences of Turkey. He has served as a member of the IEEE Fellow Committee during 2005-2008 and 2018-2020. He is currently a member of the IEEE ComSoc Awards Standing Committee between 2022–2024. He was an Editor of the IEEE TRANSACTIONS ON COMMUNICATIONS during 1995-2000.

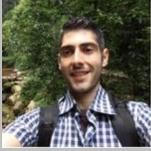
**DR. Panagiotis D. Diamantoulakis** (Senior Member, IEEE) received the Diploma (five years) and the Ph.D. degree from the Department of Electrical and Computer Engineering, Aristotle University of Thessaloniki, Thessaloniki, Greece, in 2012 and 2017, respectively. Since 2022 he is a Postdoctoral Fellow with the Department of Applied Informatics, University of Macedonia, Thessaloniki, Greece. Since 2017, he has been a Postdoctoral Fellow with Wireless Communications and Information Processing (WCIP) Group, AUTH and since 2021, he has been a Visiting Assistant Professor with the Key Lab of Information Coding and Transmission, Southwest Jiaotong University, Chengdu, China. His research interests include optimization theory and applications in wireless networks and smart grids, game theory, and optical wireless communications. He is also an Editor of IEEE Wireless Communications Letters, IEEE Open Journal of the Communications Society, Physical Communications (Elsevier), and Frontiers in Communications and Networks.

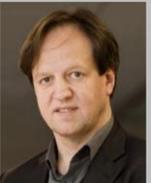
**Prof. Harald Haas** (Fellow, IEEE) received the Ph.D. degree from the University of Edinburgh, Edinburgh, U.K., in 2001. He is the Director of the LiFi Research and Development Center, University of Strathclyde, Glasgow, U.K. He is also the Initiator, the Co-Founder, and the Chief Scientific Officer of pureLiFi Ltd., Edinburgh. He has authored 550 conference and journal papers. His main research interests are in optical wireless communications, hybrid optical wireless and RF communications, spatial modulation, and interference coordination in wireless networks. Dr. Haas received the Outstanding Achievement Award from the International Solid State Lighting Alliance in 2016 and the Royal Society Wolfson Research Merit Award. His team invented spatial modulation. He introduced LiFi to the public at an invited TED Global talk in 2011. He was elected a Fellow of the Royal Society of Edinburgh in 2017. He was elected a Fellow of the Royal Academy of Engineering in 2019.

# Optical Wireless Communication Based Smart Ocean Sensor Networks for Environmental Monitoring

Ikenna Chinazaekpere Ijeh

*Department of Electrical & Electronic Engineering, Alex Ekwueme Federal University, Ndufu-Alike Ikwo, Ebonyi State, Nigeria, ikenna.ijeh@funai.edu.ng*

## I. INTRODUCTION

The demand for advanced technologies to monitor, analyze, and manage the world's oceans has led to the development of smart oceans. Smart oceans of the future will involve the development of interconnected applications through the underwater internet of things [1]. This will involve the use of smart sensing technologies such as buoys or autonomous ocean sensors, unmanned underwater vehicles for remote monitoring and data collection, and the use of machine learning for predictive modeling and forecasting, among other advancements.

The oceanic environment is increasingly vulnerable to climate change and pollution, such as gas flaring in some coastal regions of Nigeria with large oil and gas deposits. This activity generates greenhouse gases and harmful pollutants that endanger human health, marine life, and the environment. To tackle these challenges, efficient monitoring of oceans is crucial, allowing for reliable impact assessment and proposing effective mitigation or adaptation strategies. An environmentally friendly and reliable underwater communication technology is therefore required to facilitate continuous data collection and transfer from smart ocean sensor networks. Among the available options, wireless communication is the most preferable due to its flexibility, cost-effectiveness, and low environmental impacts. **Optical wireless communication (OWC)** is a most promising technology for high-speed data transmission, allowing for real-time monitoring over short-to-moderate ranges with low power consumption [2, 3].

Some environmental monitoring applications of OWC in smart ocean sensor networks include tracking changes in an ocean's temperature, salinity, dissolved oxygen, and pH levels. Other applications include monitoring carbon sequestration efforts, offshore oil and gas platforms, water quality for aquaculture operations, and early signs of natural disasters such as tsunamis or environmental threats like oil spills, to enable an effective response.

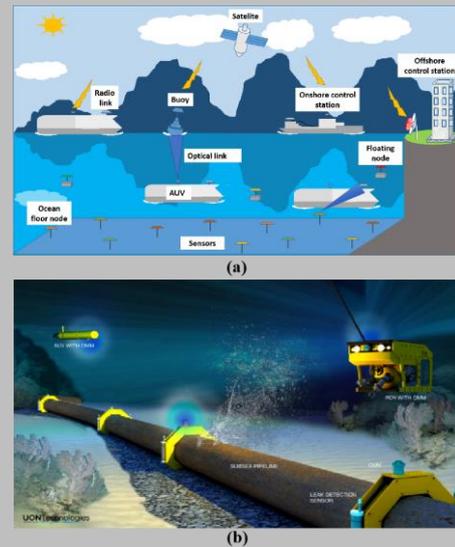

**Fig. 1.** (a) Illustration of OWC used in a smart ocean sensor network for data retrieval and transmission, and (b) An optical modem module deployed in monitoring underwater pipelines (reproduced from [4] with reference therein).

## II. CONCEPT OF OWC BASED SMART OCEAN

Fig. 1(a) shows a smart ocean sensor network for environmental monitoring, where an OWC link is used for data transfer between an autonomous underwater vehicle (AUV) and an underwater sensor node. Subsequently, the AUV can either transmit the data directly to a sea surface platform or go afloat, for the onwards transmission to a satellite. Thus, making data available for control stations to analyze and make decisions. A practical application scenario is shown in Figure 1b, where optical modem modules from UON Technologies are attached to both an underwater infrastructure and remote operated drone for monitoring and data collection [4].

**System architecture:** The smart ocean architecture could comprise of a 5-layer system which accounts for both the underwater and above-water environment (See Figure 2) [1]. It consists of a sensing, communication, networking, fusion, and application layer. In the underwater environment, the sensing layer provides identification and sensing of data using sensors and cameras. Thereafter, the communication layer handles the data transmission via a suitable communication technology (e.g. radio, acoustic, magnetic induction, optical) based on the application demand. The network layer utilizes the underwater to above water communication channel, which may involve direct or multi-hop links, to transfer data from sensor nodes and sensing devices hosted by AUVs, buoys, or other sea surface platforms to the fusion layer.

Above the water surface, the fusion layer uses cloud and edge servers to process large amounts of received data using fog and cloud computing technology. Lastly, the application layer, based on the processed data from the fusion layer, provide secured smart services for use cases such as ocean monitoring and exploration, underwater robotics, port security, etc. [1].

**Underwater OWC system:** Several organizations are committed to promoting sustainable ocean development through advanced ocean technology, and among them is the SFI Smart Ocean. Their primary focus lies in the development of autonomous and smart underwater wireless sensor networks. In the bid for efficient data exploration, the test cases of the SFI Smart Ocean pilot demonstrator included the deployment of underwater acoustic monitoring system to collect data transmitted from a rig by modems, cNODE and W-Sense. However, acoustic, and even radio frequency (RF) technologies are not capable of meeting the growing demands for high data rate and low latency underwater transmissions over a considerable link range with minimum energy requirement, all which OWC offers, see Table I.

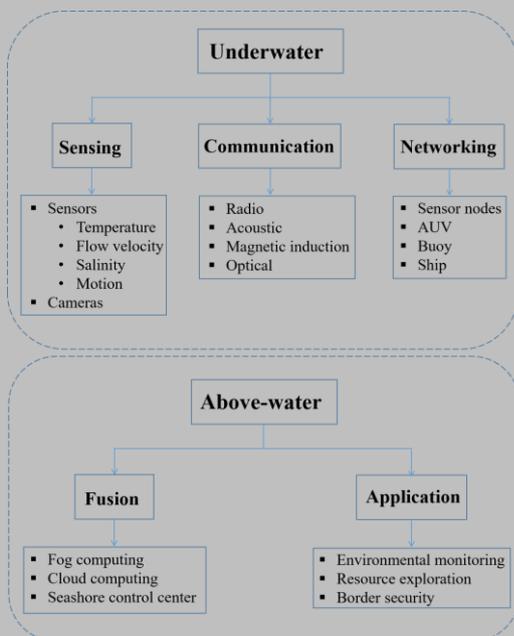

**Fig. 2.** System architecture of a smart ocean (inspired from [1]).

There are underwater OWC (UOWC) systems (consisting of a transmitter light source and a receiver photodetector) available commercially or as prototypes, suitable for various use cases. For example, Shimadzu Corporation recently developed MC500 UOWC modem, which using green and blue laser light source can achieve data rates up to 20 Mbps over a range of 80 m. Another optical modem called the LUMA X from Hydromea can achieve up to 10 Mbps across more than 40 m, while the BlueComm 200 (made up of light-emitting diodes and photomultiplier tube) from Sonardyne can operate up to 150 m and achieve a data rate of 10 Mbps. Their UV counterparts, LUMA X-UV and BlueComm 200 UV, still maintaining same data rates operates at ranges of up to 30 and 75 m respectively. These UV series make use of an ultra-violet light source, hence are resilient to the presence of other light sources.

On the other hand, prior to commercialization, some UOWC system prototypes have been developed and tested. MIT researchers developed the AquaOptical modem, which could transmit Mbps of data over a few meters. An improved version, AquaOptical II, achieved higher transmission ranges and data rates. Ifremer, Institut Fresnel research lab and Osean SAS Co. developed an optical modem based on the highly sensitive silicon photomultiplier. For internet applications, the Aqua-Fi, a portable, low-cost and energy efficient UOWC system was showcased in a laboratory setup (see [4] and references therein).

**Table 1.** Comparison of underwater wireless communication technologies (reproduced from [4] and reference therein).

| Features | RF | Acoustic | Optical |
|---|---|---|---|
| Data rate | hundreds of kbps | kbps | hundreds of Mbps |
| Bandwidth | MHz | 100 kHz | 150 MHz |
| Latency | High | Moderate | Low |
| Range | up to 10 m | hundreds of kms | up to 100 m |
| Transmit power | up to hundreds of Watts | tens of Watts | few Watts |

III. STATE OF THE ART

Recent studies on the UOWC technology are focused on improving its various aspects to make it more reliable, efficient, and cost-effective. Energy efficiency is crucial to increase the lifetime of the network. Low-power light sources and transmission protocols can be used to optimize energy consumption. Another option could be the use of sensor nodes to harvest energy from ambient underwater sources [2]. Some other considered areas include [2-5]:

**Channel characterization:** Accurately characterizing the underwater optical channel is important for optimizing

UOWC system performance. Techniques such as empirical measurements, theoretical models, and simulations are used to study these channels. Water turbidity, scattering, absorption, and turbulence can negatively impact the propagating light signals. There is currently a focus on accurate modeling of vertical transmissions as parameters such as temperature and salinity vary significantly in vertical links compared to horizontal links and can impact system performance.

**Transceiver components:** Advanced technologies for light emitting diode (LED)-based UOWC systems include power-efficient blue LEDs with phosphor conversion layers, ultraviolet (UV) LED sources, and pressure-neutral resin casting for small and light optical transceivers. However, converted green or amber LEDs result in reduced bandwidth, and UV LEDs trades off reduced impact of solar radiation to increased water absorption. Micro LED arrays and quantum-dot LEDs require further development before being used in commercial UOWC systems. Photodetectors (PDs) suitable for low-light high-speed applications include photomultiplier tubes and silicon photomultiplier, but they can be affected by solar radiations. For higher light levels in shallow waters and short ranges, positive-intrinsic-negative PDs and avalanche PDs are suitable. Single-color LEDs with large active areas can also serve as PDs for low-cost applications in special cases.

**Noise:** Experimental testing of optical underwater communication is often done in darkened laboratories to avoid interference from ambient sunlight that can affect system performance. Optical bandpass filters are commonly used to reduce sunlight interference, but a newer approach is to use a liquid crystal display as a dynamic optical filter or adaptive optical aperture to mitigate ambient light and even interference in multi-user scenarios.

**Hybrid communications:** Adopting RF communication in the underwater environment especially in seawater is very limited due to its high conductivity which increases the attenuation of the propagating electromagnetic wave hence limiting link range and achievable data rate. The conventional hybrid approach for underwater communication is the optical-acoustic system setup. For instance UOWC is used for high data rate near-range inter-autonomous underwater vehicle communication and acoustics are used for signaling events between sensor nodes or exchanging mission objectives with mission control. However, a promising and less explored area is the combination of magneto-inductive (MI) communication with UOWC. UOWC is suitable for visibly clear water scenarios and relatively unaffected by changes in salinity, but its performance deteriorates in turbid waters. On the other hand, MI is less sensitive to water turbidity but experiences significant signal attenuation in high salinity waters.

**Localization:** UOWC localization schemes can be categorized into distributed and centralized schemes. In distributed schemes, every underwater optical sensor node localizes itself by communicating with multiple anchor nodes using time of arrival and received signal strength (RSS) based localization techniques. In centralized schemes, the location information is sent periodically to the underwater optical sensor nodes by a surface buoy or sink node. Only the RSS technique has been known to be applied in centralized schemes. Commercial UOWC localization systems include Bluecomm from Sonardyne and Anglerfish from STM.

## IV. CHALLENGES AND FUTURE WORK

Data transmission via the UOWC link is challenging and depends on several factors majorly related to the aquatic channel and the practical constraints. For instance, the link range is limited by signal attenuation due to beam absorption and scattering, as well as by oceanic turbulence. In the absence of efficient localization schemes, misalignment of UOWC transceivers possibly due to ocean currents and waves will affect the link reliability. Also, solar radiations or even lights from underwater vehicles can affect the transmission quality. In addition, the limited energy capacity of the underwater sensors or devices constraints the optical transmit power and hence, limits link range and data rate.

Several research directions are necessary to tackle the challenges hindering the widespread deployment of UOWC systems. An important aspect for consideration is the underwater-to-above water communications as it ensures the accessibility of data for ease of processing and analysis. This could entail direct or relay points from water-to-air mediums with UOWC to same or mixed communication links, such as UOWC-to-terrestrial OWC or UOWC-to-RF, exploiting either the high data rate of OWC links or the easy omni-directionality and long-range transmission capability of the RF link. Relevant research in this area would involve the development of efficient relay and physical-layer security schemes.

In the underwater environment, it is important to conduct adequate investigation and accurate modeling of the UOWC channel to propose effective solutions for mitigating signal intensity degradation, particularly for vertical links, as there have been limited studies in this area. Multi-hop transmission through the use of AUV swarms and floating sensor nodes as relays could increase link range and enable energy transfer.

Furthermore, the effective and uninterrupted functioning of the increasing number of sensors can only be achieved through the implementation of wireless power transfer technology. Localization, especially for node tracking in the dynamic underwater environment, is a critical aspect of UOWC systems. It is important to consider the impact of transmission quality, range, and energy on localization performance. Moreover, it is necessary to shift considerable focus towards higher layer aspects of UOWC systems, such as developing efficient routing protocols and network architectures to enhance connectivity and coverage [2, 5]. Lastly, the use of artificial intelligence should be explored to develop intelligent and self-learning functionality for smart ocean sensor networks in the constantly changing underwater environment.

REFERNCES

[1] T. Qiu, Z. Zhao, T. Zhang, C. Chen and C. L. P. Chen, "Underwater internet of things in smart ocean: system architecture and open issues," in *IEEE Transactions on Industrial Informatics*, vol. 16, no. 7, pp. 4297-4307, July 2020.

[2] N. Saeed, A. Celik, T. Y. Al-Naffouri, and M. S. Alouini, "Underwater optical wireless communications, networking, and localization: A survey", in *Ad Hoc Networks*, vol. 94, p. 101935, Nov. 2019.

[3] M. A. Khalighi, C. J. Gabriel, L. M. Pessoa, and B. Silva, "Underwater visible light communications, channel modeling and system design," in *Visible Light Communications: Theory and Applications*. Boca Raton, FL, USA: CRC Press, 2017, pp. 337–372.

[4] I. C. Ijeh, "Investigation of random channel effects on the performance of underwater wireless optical communication links," in *Ph.D. Thesis*, École Centrale Marseille, Dec. 2021.

[5] P. A. Hoeher, J. Sticklus and A. Harlakin, "Underwater optical wireless communications in swarm robotics: A tutorial," in *IEEE Communications Surveys & Tutorials*, vol. 23, no. 4, pp. 2630-2659, Fourthquarter 2021.

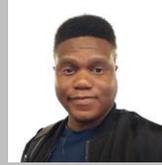
**Dr. Ikenna Chinazaekpere Ijeh** (Member; IEEE, COREN, NSE) is from Abakaliki, Ebonyi State, southeastern Nigeria. He received a B.Eng. degree in Electrical Electronics Engineering from Caritas University, Enugu, Nigeria, in 2013, an M.Sc. degree in Control and Instrumentation from the University of Derby, United Kingdom, in 2016, and a Ph.D. degree in Optics, Photonics and Image Processing from École Centrale de Marseille, France, in 2021. He is currently a Researcher and Lecturer with the Department of Electrical & Electronic Engineering, Alex-Ekwueme Federal University Ndufu-Alike Ikwo, Ebonyi State, Nigeria. He is a Working Group Member of H2020 COST Action CA19111 - "European Network on Future Generation Optical Wireless Communication Technologies (NEWFOCUS)". Amongst his full scholarship awards for M.Sc. and Ph.D. programmes, he was awarded bourse d'études pour les doctorants étrangers de la Ville de Marseille in 2020. His research interests include wireless communication systems and the automation of control systems.